\documentclass[journal]{IEEEtran}

\usepackage{lipsum}
\usepackage{graphicx}
\ifCLASSOPTIONcompsoc
    \usepackage[caption=false, font=normalsize, labelfont=sf, textfont=sf]{subfig}
\else
\usepackage[caption=false, font=footnotesize]{subfig}
\fi

\usepackage{graphicx}
\usepackage{amsfonts}
\usepackage{amsmath}
\usepackage{bm}
\usepackage{cite}
\usepackage{epstopdf}
\usepackage{booktabs}
\usepackage{lipsum}
\usepackage{threeparttable}
\usepackage{balance}
\usepackage{amssymb}
\usepackage{ntheorem}
\usepackage[ruled,linesnumbered]{algorithm2e}
\usepackage{multirow}
\usepackage{framed}
\usepackage{color}
\usepackage{float}

\theorembodyfont{\rmfamily}

\newtheorem{fact}{Fact}

\begin{document}
	
\title{Two-Dimensional DOA Estimation for L-shaped Nested Array via Tensor Modeling}

\author{Feng Xu, \IEEEmembership{Member, IEEE}, and Sergiy A. Vorobyov, \IEEEmembership{Fellow, IEEE}
\thanks{This work was supported in part by the Academy of Finland under Grant 319822 and in part by the China Scholarship Council. This work was conducted while Feng Xu was a visiting doctoral student with the Department of Signal Processing and Acoustics, Aalto University. (\textit{Corresponding author: Sergiy A. Vorobyov.})}
\thanks{Feng Xu was with the School of Information and Electronics, Beijing Institute of Technology. He also was and he is with the Department of Signal Processing and Acoustics, Aalto University. (e-mail: feng.xu@aalto.fi).}
\thanks{Sergiy A. Vorobyov is with the Department of Signal Processing and Acoustics, Aalto University, Espoo 02150, Finland. (e-mail: svor@ieee.org).}
}
\maketitle

\begin{abstract}
The problem of two-dimensional (2-D) direction-of-arrival (DOA) estimation for the L-shaped nested array is considered. Typically, the multi-dimensional structure of the received signal in co-array domain is ignored in the problem considered. Moreover, the cross term generated by the correlated signal and noise components degrades the 2-D DOA estimation performance seriously. To tackle these issues, an iterative 2-D DOA estimation approach based on tensor modeling is proposed. To develop such approach, a higher-order tensor is constructed, whose factor matrices contain the sources azimuth and elevation information. By exploiting the Vandermonde structure of the factor matrix, a computationally efficient tensor decomposition method is then developed to estimate the sources DOA information in each dimension independently. The pair-matching of the azimuth and elevation angles is conducted via the cross-correlation matrix (CCM) of the received signals. An iterative method is further designed to improve the DOA estimation performance. Specifically, the cross term is estimated and removed in the next step of such iterative procedure on the basis of the DOA estimates originated from the tensor decomposition in the previous step. As a consequence, the DOA estimation with better accuracy and higher resolution is obtained. The proposed iterative 2-D DOA estimation method for the L-shaped nested array can resolve more sources than the number of real elements, which is superior to conventional approaches. Simulation results validate the performance improvement of the proposed 2-D DOA estimation method as compared to existing state-of-the-art DOA estimation techniques for the L-shaped nested array.
\end{abstract}


\section{Introduction}
\IEEEPARstart{T}{he} problem of two-dimensional (2-D) direction-of-arrival (DOA) estimation of multiple signals impinging on an antenna array has attracted considerable attention in several applications such as wireless communications, radar, sonar and others \cite{1,2,3}. In these applications, several 2-D array structures such as uniform rectangular array (URA), circular array and L-shaped array have been considered \cite{33}. For example, URA is a widely used array geometry for airborne or spaceborne arrays, where DOA estimation methods such as multiple signal classification (MUSIC) \cite{4} and estimation of signal parameters via rotational invariance technique (ESPRIT) \cite{5,6} can be straightforwardly conducted. However, it has been established that L-shaped array is superior to URA \cite{7} since the corresponding Carmer-Rao bound (CRB) is significantly (37\%) lower than that of the URA. This property means that L-shaped array can provide a higher accuracy for 2-D DOA estimation. Therefore, the study of high resolution 2-D DOA estimation methods for L-shaped array has been the focus of array processing research over the past two decades \cite{8,31,9,10,11,12,13,32,14,15,16,17,18,34,19}.

In general, L-shaped array can be divided into two linear subarrays. Thanks to it, the 2-D DOA estimation problem can be regarded as two one-dimensional (1-D) DOA estimation problems, and 1-D high resolution DOA estimation methods can be generalized to the case of L-shaped array conveniently \cite{8,31,9,10,11}. For example, the MUSIC algorithm for solving two 1-D spectrum searching problems is much easier than conducting a complex 2-D spectrum search. To further reduce the computational complexity, a modified propagator method (PM) has been proposed \cite{8}, which avoids the use of matrix singular value decomposition (SVD). In \cite{31}, the azimuth and elevation angles are independently estimated using a 1-D subspace-based method without eigendecomposition. The study of DOA estimation in the presence of mutual coupling has also been considered for {  the} L-shaped array \cite{9}. Note that the independent sets of azimuth and elevation angles must be properly paired after using the 1-D DOA estimation methods. Several approaches have been developed in the literature for this purpose \cite{10,11}. To obtain the correct azimuth and elevation pairs, a Toeplitz matrix has been built by exploiting the cross-correlation matrix (CCM) of the signal received by {  the} L-shaped array in \cite{10}. The common structure of the above reviewed approaches for L-shaped array-based 2-D DOA estimation consists of two parts: 1) 1-D DOA estimation for each subarray, and 2) a proper pair-matching of the two estimated angle sets. Let us categorize these approaches in the first category.

The second category of approaches for L-shaped array-based 2-D DOA estimation \cite{12,13,32,14,15,16,17,18,34,19}, on the contrary, aims at achieving an automatic pairing during the joint 2-D DOA estimation. By exploiting the fact that the noise component can be fully eliminated in the CCM, a joint SVD method \cite{12,13,32} has been proposed to improve the 2-D DOA estimation performance. The auto-pairing of two estimated angle sets is then also achieved by using the eigenvalue decomposition (EVD) of the submatrices constructed from the CCM. In \cite{14}, an angle estimation method has been introduced to split the joint steering vector into two steering vectors in order to fulfill the L-shaped DOA estimation without pairing. However, this method requires one 1-D spectrum search and suffers from the angle ambiguity. To reduce the computational burden, a signal subspace-based algorithm has been proposed in \cite{15}. This method estimates the noise subspace by rearranging the elements of three vectors, i.e., the first column, the first row and the diagonal entries of the CCM. A generalization of the PM using the CCM has been developed in \cite{16}, where only linear operations on the signal matrix have been required. Furthermore, the conjugate symmetry property of the array manifold has also been exploited in \cite{17,18,34} to raise the array aperture and the number of snapshots simultaneously. Recently, a tensor-based approach \cite{19} aiming at increasing the {   system's degrees of freedom} (DOF) has been suggested for {  the} L-shaped array-based DOA estimation, where the subarrays on both axes are divided into several overlapping subarrays of smaller size. A more general case has been studied in \cite{20}, where the authors illustrate that any centrosymmetric array processing problem can be interpreted in terms of a coupled canonical polyadic decomposition (CPD) problem. By using tensor modeling, the multi-dimensional structure of the received signal is exploited and, hence, the estimation performance can be improved \cite{35,40}.

Meanwhile, a special array geometry named as nested array has been widely investigated \cite{21,22}, largely thanks to the fact that it can be used to detect more sources than the number of real antenna elements due to the increased DOF in co-array domain.\footnote{Co-array of a given array is the array whose elements are located at the positions determined by the differences between physical sensor locations (see \cite{21,22} and the references therein).} When uniform linear array (ULA) is replaced by linear nested array for {  the} L-shaped array, the corresponding array is named as {\it L-shaped nested array} \cite{23,24,25,26,27}. {  The} L-shaped nested array also may enjoy an increased DOF. In \cite{23,27}, the authors consider the 2-D DOA estimation problem as two independent 1-D DOA estimation problems. For each 1-D problem, the spatial smoothing (SS) (see \cite{21}) is used directly, while the fourth-order difference co-array is exploited in \cite{24}. The azimuth and elevation angles are matched by pairing the source powers estimated by two nested subarrays separately. The signal subspace joint diagonalisation (SSJD) technique \cite{25} is used to conduct 2-D DOA estimation and pair-matching, which can be regarded as a generalization of \cite{17,18}. Possible holes in the cross-difference co-array for {  the} L-shaped nested array can be filled by using oblique projection \cite{26}, and a virtual CCM with larger aperture can be constructed to fulfill 2-D DOA estimation with better performance.

It is important to stress, however, that the existing methods for {  the} L-shaped nested array DOA estimation ignore the multi-dimensional structure of the received signal, especially after SS is applied on both directions. Besides, in the co-array domain, the signal and noise terms become correlated and the cross term between them cannot be ignored. This unexpected component can degrade the DOA estimation performance significantly. To tackle these problems, in this paper, an iterative 2-D DOA estimation method via tensor modeling is developed for {  the} L-shaped nested array. The contributions of this paper are the following.
\begin{itemize}
    \item Unlike conventional techniques for {  the} L-shaped nested array which average the received signals of all subarrays in co-array domain by applying SS, a higher-order tensor model is designed here to store those received signals in order to exploit the multi-dimensional structure inherent in the signals. The cross term between the correlated signal and noise terms is also considered. {  The three components of the designed tensor model, i.e., the signal component, cross term component and noise component, are explicitly derived. The parameter identifiability of the designed tensor model is also studied. Based on this study, the number of subarrays for SS is optimized to maximize the system DOF.}
    \item A computationally efficient tensor decomposition method is proposed for 2-D DOA estimation. The azimuth and elevation angles are paired by the joint sources spatial information in the CCM of the received signals.
    \item An iterative DOA estimation method for {  the} L-shaped nested array is proposed. The essence of the iterative method is that the cross term is estimated and removed in the next step based on the DOA estimates obtained at the previous step via tensor decomposition. The estimated received signal is then modified and can be used as an input for more accurate DOA estimation. Thus, the DOA estimation performance can be improved gradually over iterations.
    \item Analytical expression for CRB associated with our proposed received signal model is derived.
\end{itemize}

The remainder of this paper is organized as follows. Some preliminaries about tensors and signal model for {  the} L-shaped nested array are introduced in Section~\ref{sec2}. A novel higher-order tensor model for the signal received by {  the} L-shaped nested array is developed in Section~\ref{sec3}. The parameter identifiability for this tensor model is studied in the same section with the purpose to demonstrate advantages of the proposed model, while the optimization of DOF is also presented. In Section~\ref{sec4}, an iterative 2-D DOA estimation method for {  the} L-shaped nested array is proposed via decomposition of the designed signal tensor. Section~\ref{sec4} also analyzes the computational complexity of the proposed method and derives the CRB of the designed signal model. Some discussions on snapshot deficient and unknown sources number scenarios are also presented. Numerical results are presented in Section~\ref{sec5} in order to verify the effectiveness of the proposed method. Finally, Section~\ref{sec6} draws our conclusion.

\textsl{Notation}: Scalars, vectors, matrices and tensors are represented by lower-case, boldface lower-case, boldface upper-case, and calligraphic letters, e.g., $r$, $\bf r$, $\bf R$, and $\cal R$, respectively. The transposition, Hermitian transposition, inversion, pseudo-inversion, conjugation, outer product, Kronecker product and Khatri-Rao (KR) product operations are denoted by ${\left( \cdot  \right)^T},{\left(  \cdot  \right)^H},{\left(  \cdot  \right)^{ - 1}}, {\left( \cdot  \right)^{\dag}}, {  {\left(\cdot\right)^*}} , \circ ,\otimes$, and $\odot$, respectively, while the operator ${\rm vec}\left\{ \cdot \right\}$ stacks the elements of a matrix/tensor one by one to a long vector. The operation denoted as ${\rm diag}({\bf{r}})$ returns a diagonal matrix built out of its vector argument, while $\left\| {\bf{R}} \right\|_{\rm F}$ and $\left\| {\bf{R}} \right\|$ stand for the Frobenius norm and Euclidean norm of ${\bf{R}}$, respectively. The operator ${\rm E}\left\{ \cdot \right\}$ is the mathematical expectation, while ${\rm tr} (\cdot)$ is the trace of a matrix. {  The notation $\triangleq$ means ``equals by definition''.} Moreover, ${{\bf{1}}_{M\times N}}$ and ${{\bf{0}}_{M\times N}}$ denote the all-one matrix of dimension $M \times N$ and the all-zero matrix of size $M \times N$, respectively, while ${{\bf{I}}_{M}}$ and ${\bf J}_M$ stands for the $M \times M$ identity matrix and the $M \times M$ exchange matrix with ones on the anti-diagonal and zeros elsewhere, respectively. For ${\bf R} \in {{\mathbb{C}}^{M \times N}}$, the $n$-th column vector and $(m,n)$-th element are denoted by ${\bf r}_n$ and $r_{mn}$, respectively, while the $m$-th element of ${\bf r} \in {{\mathbb{C}^{M}}}$ is given by $r_m$. The estimates of $\bf R$, $\bf r$ {  and $r$} are denoted as $\bf \hat R$, $\bf \hat r$  {  and $\hat r$}, respectively, while the noise-free versions of $\bf R$, $\bf r$  {  and $r$} are written as $\bf \tilde R$, $\bf \tilde r$  {  and $\tilde r$, respectively. The $(i_1,\cdots, i_N)$-th element of $N$-order tensor ${\cal R}$ is denoted as $[{\cal R}]_{i_1...i_N}$.}

\section{Preliminaries on Tensors and Signal Model}\label{sec2}
In this section, we introduce some preliminaries about tensor \cite{28,29,30}, which will be heavily used later in the paper. Then, the signal model for {  the} L-shaped nested array is given.

\subsection{Preliminaries on Tensors}
\begin{fact}(Kruskal Form Tensor):
An $N$-th order tensor ${{\cal R} \in {{\mathbb{C}}^{{I_1} \times {I_2}\times \cdots \times {I_N}}}}$ is presented in Kruskal form if
\begin{equation} \label{KruskalForm}
\begin{footnotesize}
\begin{aligned}
{\cal R}  =  \sum\limits_{k = 1}^K {{\lambda}_k{{\bm{\alpha}}_k^{(1)}} \circ {{\bm{\alpha}}_k^{(2)}} \circ \cdots \circ {{\bm{\alpha}}_k^{(N)}}}\triangleq  \left[\left[{\bm \lambda}; {\bf A}^{(1)}, {\bf A}^{(2)},\cdots, {\bf A}^{(N)}\right]\right],
\end{aligned}
\end{footnotesize}
\end{equation}
where ${\bm \lambda} \triangleq [\lambda_1,\lambda_2,\cdots, \lambda_K]^T$, ${{\bm{\alpha}}_k^{(n)}}$ is the $k$-th column of ${\bf{A}}^{(n)}$ with ${\bf{A}}^{(n)} \in {\mathbb{C}^{I_n \times K}}$ being the $n$-th factor matrix, and $K$ is the tensor rank. Following this type of tensor presentation, the CPD of a rank $K$ tensor $\cal R$ consists of finding ${\cal \hat R} = \left[\left[{\bm \lambda}; {\bf A}^{(1)}, {\bf A}^{(2)},\cdots, {\bf A}^{(N)}\right]\right]$ so that $\left|\left|{\cal R}-{\cal \hat R}\right|\right|_{\rm F}^2$ is minimized.
\end{fact}

\begin{fact}(Tensor Reshape):
For an $N$-th order tensor ${{\cal R} \in {{\mathbb{C}}^{{I_1} \times {I_2}\times \cdots \times {I_N}}}}$, the tensor reshape operator generates an $M$-th order tensor ${{\cal T} \in {{\mathbb{C}}^{{J_1} \times {J_2}\times \cdots \times {J_M}}}}$ that satisfies ${\rm vec}\{{\cal R}\}  = {\rm vec}\{{\cal T}\}$ and ${\prod_{n=1}^{N}}I_n = {\prod_{m=1}^{M}}J_m$. In particular, consider the set ${\mathbb{A}} = \{1,2,\cdots,N\}$ and $M$ subsets ${\mathbb{A}}_m$, $m = 1,2,\cdots,M$, as partitions of ${\mathbb{A}}$ such that ${\mathbb{A}}_1 \cup \cdots \cup{\mathbb{A}}_M = {\mathbb{A}}$ and ${\mathbb{A}}_i \cap {\mathbb{A}}_j = \emptyset, \forall i \neq j$. Then, the tensor reshape operator can be expressed as
\begin{equation} \label{Reshape}
  {\cal T} \triangleq {\rm reshape} ({\cal R}, [{\mathbb{A}}_1,{\mathbb{A}}_2,\cdots,{\mathbb{A}}_M]).
\end{equation}

For example, taking a 5-th order tensor ${{\cal R} \in {{\mathbb{C}}^{{I_1} \times {I_2}\times \cdots \times {I_5}}}}$, let ${\mathbb{A}}_1 = \{3,1\}$, ${\mathbb{A}}_2 = \{4,2\}$ and ${\mathbb{A}}_3 = \{5\}$. The reshaped 3-rd order tensor ${\cal T}$ is of dimension ${{\mathbb{C}}^{{I_1I_3} \times {I_2 I_4}\times {I_5}}}$, i.e., $J_1 = I_1I_3, J_2 = I_2I_4$, and $J_3 = I_5$.
\end{fact}

\begin{fact}(Tensor Unfolding):
For an $N$-th order tensor ${\cal R} = \left[\left[{\bm \lambda}; {\bf A}^{(1)}, {\bf A}^{(2)},\cdots, {\bf A}^{(N)}\right]\right]$ and ${\bm \Lambda} = { {\rm diag}}({\bm \lambda})$, the unfolding of ${\cal R}$ from the $n$-th dimension returns a matrix ${\bf R}_{(n)} \in {\mathbb{C}^{I_1\cdots I_{n-1}I_{n+1}\cdots I_N \times I_n}}$ such that
\begin{footnotesize}
\begin{equation} \label{Unfolding}
  {\bf R}_{(n)} = \left( {{{\bf{A}}^{(N)}} \cdots \odot {{\bf{A}}^{(n + 1)}} \odot {{\bf{A}}^{(n - 1)}} \cdots \odot{{\bf{A}}^{(1)}}} \right){\bm \Lambda}\left({{{{\bf{A}}^{(n)}}}}\right)^T.
\end{equation}
\end{footnotesize}

It is worth noting that tensor unfolding can be regarded as a special case of tensor reshape where only two subsets ${\mathbb{B}}_1 = \{1,2,\cdots,n-1,n+1,\cdots, N\}$ and ${\mathbb{B}}_2  = \{n\}$ exist. Hence, we define the unfolding operator as ${\bf R}_{(n)} \triangleq {\rm unfolding} ({\cal R},[{\mathbb{B}}_1,{\mathbb{B}}_2])$.
\end{fact}

{  \begin{fact}(Cross-correlation Tensor of Two Matrices):
	For two matrices ${\bf A} \in {\mathbb{C}^{I_1 \times I_2}}$ and ${\bf B} \in {\mathbb{C}^{J_1 \times J_2}}$, the cross-correlation tensor ${\cal R}$ is a 4-order tensor of size $I_1 \times I_2 \times J_1 \times J_2$ \cite{47}, whose $(i_1,i_2,j_1,j_2)$-th element is given by
	\begin{equation}\label{eq4}
		 [{\cal R}]_{i_1i_2j_1j_2} = a_{i_1i_2}b_{j_1j_2}^*.
	\end{equation}
\end{fact}}

\subsection{Signal Model}
\begin{figure}
\centerline{\includegraphics[width=0.8\columnwidth]{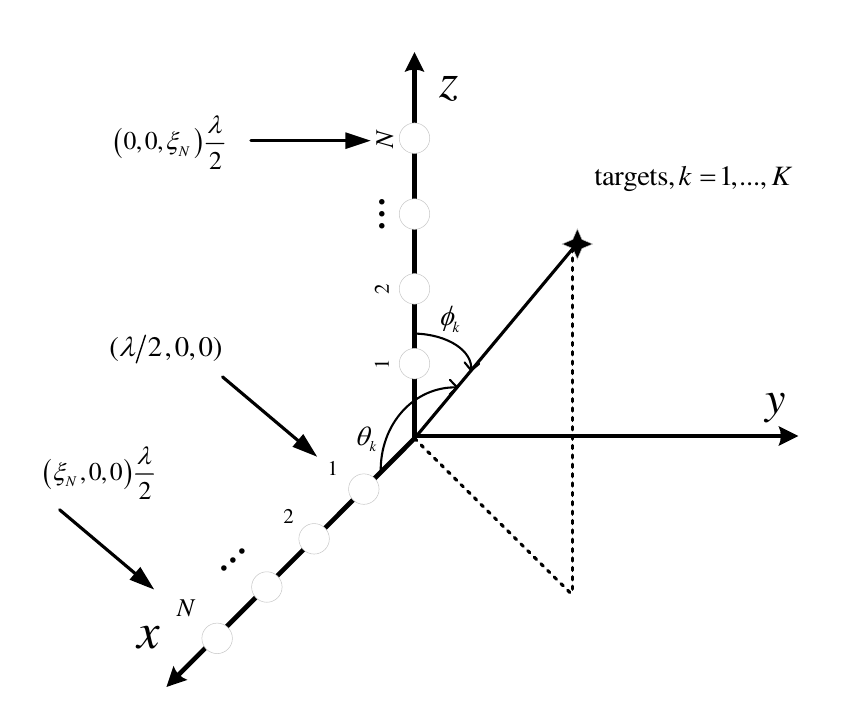}}
\caption{{  L-shaped nested array configuration with $2N$ physical sensors.}}\label{sys}
\end{figure}

Consider an L-shaped nested array that consists of two linear subarrays arranged along the x-axis and z-axis, as shown in Fig.~1. Each subarray is a two-level nested array with $N$ ($N$ is even) elements. The first level has $N/2$ elements with spacing $d_1$ and the second level has another $N/2$ elements with spacing $d_2$. Without loss of generality, let $d_1 = \lambda/2$ and $d_2 = (N/2+1)d_1$\cite{21}, where $\lambda$ is the working signal wavelength. {  The coordinates of the elements on x-axis and z-axis subarrays are $\left( \{\xi_n\}_{n=1}^N \lambda/2, 0,0 \right)$ and $\left( 0, 0, \{\xi_n\}_{n=1}^N \lambda/2 \right)$, respectively, where $\{\xi_n\}_{n=1}^N = \{1, 2, \cdots, N/2, N/2+1, N+2, \cdots, N/2(N/2+1)\}$.} Assume that $K$ spatial-temporal uncorrelated narrowband far-field sources are impinging on the L-shaped nested array with azimuth and elevation angles $\{(\theta_k,\phi_k)\}_{k=1}^K$. The steering vectors of both subarrays are given as ${\bf a}_{\rm x} (\theta) \triangleq [e^{-j \pi \cos \theta \xi_1}, e^{-j \pi \cos \theta \xi_2}, \cdots, e^{-j\pi\cos\theta\xi_{N}}]^T \in {\mathbb{C}^N}$ and ${\bf a}_{\rm z} (\phi) \triangleq [e^{-j\pi\cos\phi\xi_1}, e^{-j\pi\cos\phi\xi_2}, \cdots, e^{-j\pi\cos\phi\xi_{N}}]^T \in {\mathbb{C}^N}$, respectively, where $\xi_n$ is the $n$-th element of the set $\{\xi_n\}_{n=1}^N$. Accordingly, the $N \times K$ steering matrices can be written as
\begin{equation}\label{Steer}
	\begin{aligned}
		& {\bf A}_{\rm x} \triangleq [{\bf a}_{\rm x} (\theta_1),{\bf a}_{\rm x} (\theta_2), \cdots, {\bf a}_{\rm x} (\theta_K)] \\
		& {\bf A}_{\rm z} \triangleq [{\bf a}_{\rm z} (\phi_1),{\bf a}_{\rm z} (\phi_2), \cdots, {\bf a}_{\rm z} (\phi_K)],
	\end{aligned}
\end{equation}
respectively.

Then, the received signal of the L-shaped nested array at the time instance $t$ can be expressed as
\begin{equation}\label{1}
    \begin{aligned}
        & {\bf x}(t) = {\bf A}_{\rm x} {\bf s}(t)+ {\bf n}_{\rm x} (t)\\
        & {\bf z}(t) = {\bf A}_{\rm z} {\bf s}(t)+ {\bf n}_{\rm z} (t),
    \end{aligned}
\end{equation}
where $t = 1, 2, \cdots, T_{\rm s}$ with $T_{\rm s}$ being the signal time duration (sample size), i.e., the number of snapshots after sampling, ${\bf s} (t) \triangleq [s_1(t), s_2(t), \cdots, s_K(t)]^T \in {\mathbb{C}^K}$ is the signal vector, ${\bf n}_{\rm x} (t)$ and ${\bf n}_{\rm z} (t)$ are the additional Gaussian white noise vectors on x-axis and z-axis, respectively.

Taking first ${\bf x}(t)$ in \eqref{1}, the auto-correlation matrix of the received signal can be expressed as ${\bf R}_{\rm x} = {\rm E}\{{\bf x} (t){\bf x}^H(t)\} = {\bf A}_{\rm x} {\bf R}_{\rm s} {\bf A}_{\rm x}^H + \sigma_{\rm n}^2{\bf I}_N$, where ${\bf R}_{\rm s} = {\rm diag}({\bf p})$ is the auto-correlation matrix of the source signals, ${\bf p} \triangleq [\sigma_1^2, \sigma_2^2, \cdots, \sigma_K^2]^T \in {\mathbb{C}^K}$, $\sigma_k^2$ denotes the power of the $k$-th source, and $\sigma_{\rm n}^2$ represents the power of the noise{\footnote{    In practice, ${\bf R}_{\rm x}$ can be estimated as the sample averaging of the received signal ${\bf x}(t)$. More discussions about estimating ${\bf R}_{\rm x}$ can be found in Subsection~\ref{subsection1}.}}. Vectorizing ${\bf R}_{\rm x}$, we have ${\bf y}_{\rm x} \triangleq {\rm vec}\left\{{\bf R}_{\rm x}\right\} = ({\bf A}_{\rm x}^* \odot {\bf A}_{\rm x}){\bf p} + \sigma_{\rm n}^2{\bf e}$, where ${\bf e}$ is the result of the identity matrix vectorization, i.e., ${\bf e} \triangleq {\rm vec} \{{\bf I}_{N}\}$. The term ${\bf A}_{\rm x}^* \odot {\bf A}_{\rm x}$ can be regarded as the steering matrix of a virtual ULA with larger aperture and more elements in co-array domain. {  In terms of signal-to-noise ratio (SNR), ${\bf y}_{\rm x}$ can be rewritten in normalized form as
\begin{equation} \label{2}
{\bf y}_{\rm x} = ({\bf A}_{\rm x}^* \odot {\bf A}_{\rm x}){\bf p} + {\bf e},
\end{equation}
where ${\bf p} \triangleq 1/\sigma_{\rm n}^2 [\sigma_1^2, \sigma_2^2, \cdots, \sigma_K^2]^T \in {\mathbb{C}^K}$ becomes the vector of source SNRs, for which we use the same notation for simplicity.}

After removing the repeated rows in ${\bf y}_{\rm x}$ and sorting the other remaining rows, the received signal can be updated as
\begin{equation}\label{3}
    {\ddot {\bf y}}_{\rm x} = {\ddot{\bf A}}_{\rm x} {\bf p} + {\ddot{\bf e}},
\end{equation}
where ${\ddot{\bf A}}_{\rm x} \triangleq \left[{\ddot{\bf a}}_{\rm x} (\theta_1), {\ddot{\bf a}}_{\rm x} (\theta_2), \cdots, {\ddot{\bf a}}_{\rm x} (\theta_K)\right] \in {\mathbb{C}^{(2S-1) \times K}}$ is the difference co-array steering matrix, ${\ddot{\bf a}}_{\rm x} (\theta) \triangleq \left[e^{-j \pi (-S+2) \cos \theta}, \cdots, e^{-j \pi \cos \theta}, \cdots,e^{-j\pi S \cos\theta}\right]^T \in {\mathbb{C}^{2S-1}}$ is the difference co-array steering vector in direction $\theta$, ${\ddot{\bf e}} \triangleq \left[{\bf 0}_{(S-1)\times 1}, 1, {\bf 0}_{(S-1)\times 1}\right]$, and $S = N/2(N/2+1)$.

The received signal in co-array domain for z-axis can be analogously expressed by replacing $\rm x$ and $\theta$ in \eqref{3} with $\rm z$ and $\phi$, respectively.

\section{Proposed Higher-Order Tensor Model for L-shaped Nested Array}\label{sec3}
The signal model \eqref{3} has only one snapshot. Like in the linear nested array case \cite{21}, the SS technique can be introduced to increase the number of snapshots. In particular, the number of subarrays $M$ and the number of elements within one single subarray $Q$ can be selected both equal to $S$ such that the average of the signal covariance matrices for all subarrays is expressed by the square of the signal covariance matrix for a ULA subarray. However, the multi-dimensional structure of the received signal in co-array domain is lost after applying the SS technique because of the averaging of the signal covariance matrices for all subarrays. The sources azimuth and elevation pairing information contained in the CCM is also ignored. Because of these problems, the DOA estimation performance for {  the} L-shaped nested array degrades.

To tackle the above mentioned problems, a higher-order tensor signal model is designed in this section in order to be able to exploit the multi-dimensional structure of the received signal for all subarrays instead of averaging them as in the SS technique. 

\subsection{High-Order Cross-Correlation Tensor Modeling}
In general, the difference co-array can be divided into $M$ overlapping subarrays, each of them containing { $Q = 2S-M$} elements. For $m$-th subarray (here $m = 1,2,\cdots,M$), the received signal in x-axis can be expressed as
\begin{equation}\label{4}
    {\ddot {\bf y}}_{\rm x}^{(m)} = {\ddot {\bf A}}_{\rm x}^{(m)}{\bf p} + {\ddot {\bf e}}^{(m)},
\end{equation}
where ${\ddot {\bf A}}_{\rm x}^{(m)} \in {\mathbb{C}^{Q \times K}}$ is the steering matrix of the $m$-th subarray and ${\ddot {\bf e}}^{(m)} \in {\mathbb{C}^Q}$ contains the elements from $m$-th to $(m+Q)$-th row of ${\ddot {\bf e}}$. Since the difference co-array is a ULA, the steering matrix for each subarray is a Vandermonde matrix, which satisfies the property that
\begin{equation}\label{5}
    {\ddot {\bf A}}_{\rm x}^{(m)} = {\dot {\bf A}}_{\rm x}{\bm \Omega}_{\rm x}^{(m)},
\end{equation}
where ${\dot {\bf A}}_{\rm x} \in {\mathbb{C}^{Q \times K}}$ is the reference steering matrix corresponding to the submatrix of ${\ddot {\bf A}}_{\rm x}$ with first $Q$ rows, ${\bm \Omega}_{\rm x}^{(m)} = {\rm diag}({\bm \kappa}_{\rm x}^{m-1})$, and ${\bm \kappa}_{\rm x} \triangleq \left[e^{-j\pi\cos\theta_1}, \cdots, e^{-j\pi\cos\theta_K}\right]^T \in {\mathbb{C}^{K}}$.

Then, let ${\bf X} \triangleq \left[{\ddot {\bf y}}_{\rm x}^{(1)},{\ddot {\bf y}}_{\rm x}^{(2)}, \cdots, {\ddot {\bf y}}_{\rm x}^{(M)}\right] \in {\mathbb{C}^{Q \times M}}$ denote the concatenation of the received signals for all $M$ subarrays. In compact form, ${\bf X}$ can be written as
\begin{equation}\label{6}
        {\bf X} = {\dot {\bf A}}_{\rm x}{\bf R}_{\rm s} {\bf K}_{\rm x}^T + {\bf W}= {\sum_{k=1}^K p_k{\dot {\bf a}}_{\rm x}(\theta_k) {\bf k}_{\rm x}^T(\theta_k)}+{\bf W},
\end{equation}
where ${\bf K}_{\rm x} \triangleq \left[{\bm \kappa}_{\rm x}^{0}, {\bm \kappa}_{\rm x}^1,\cdots,{\bm \kappa}_{\rm x}^{M-1}\right]^T \in {\mathbb{C}^{M \times K}}$ denotes the phase rotation matrix between different subarrays, ${\dot {\bf a}}_{\rm x}^{(1)}(\theta_k) \triangleq \left[e^{-j\pi (-S+2)\cos\theta_k}, \cdots, e^{-j\pi (-S+Q+1)\cos\theta_k}\right]^T \in {\mathbb{C}^{Q}}$ and ${\bf k}_{\rm x} (\theta_k) \triangleq \left[1, e^{-j\pi\cos\theta_k}, \cdots, e^{-j\pi (M-1) \cos \theta_k}\right]^T \in {\mathbb{C}^{M}}$ are the $k$-th columns of the matrices ${\dot {\bf A}}_{\rm x}$ and ${\bf K}_{\rm x}$, respectively, and ${\bf W} \triangleq \left[{\ddot {\bf e}}^{(1)}, {\ddot {\bf e}}^{(2)}, \cdots, {\ddot {\bf e}}^{(M)} \right] \in {\mathbb{C}^{Q \times M}}$ is the noise matrix, which can also be expressed as
\begin{equation}\label{7}
{\bf{W}} = \left\{ \begin{array}{l}
{\left[ {{{\bf{0}}_{M \times (S - M)}},{  {{\bf{J}}_M}},{{\bf{0}}_{M \times (S - M)}}} \right]^T},\quad M \le S\\
\\
\left[ {{{\bf{0}}_{Q \times (S - {\rm{Q}})}},{  {{\bf{J}}_Q}},{{\bf{0}}_{Q \times (S - Q)}}} \right],\quad M > S .
\end{array} \right.
\end{equation}
It can be observed that the rank of the noise matrix ${\bf W}$ is equal to $\min \{M,Q\}$.

Similarly, the matrix form of the received signal for all subarrays in z-axis can be modeled as
\begin{equation}\label{8}
	 {\bf Z} = {\dot {\bf A}}_{\rm z}{\bf R}_{\rm s} {\bf K}_{\rm z}^T + {\bf W}= {\sum_{k=1}^K p_k {\dot {\bf a}}_{\rm z}(\phi_k) {\bf k}_{\rm z}^T(\phi_k)} + {\bf W},
\end{equation}
where ${\dot {\bf A}}_{\rm z}$, ${\bf K}_{\rm z}$, ${\dot {\bf a}}_{\rm z}^{(1)}(\phi_k)$ and ${\bf k}_{\rm z}(\phi_k)$ are defined analogously to their counterparts in x-axis.

To fully exploit the multi-dimensional structure of the received signals for the subarrays in x-axis and z-axis, {  the cross-correlation tensor of the received signal matrices ${\bf X}$ and ${\bf Z}$ needs to be considered. To establish the relationship between the designed cross-correlation tensor ${\cal R}_{\rm xz}$ and ${\bf X} \otimes {\bf Z}^*$ let us first focus on the construction of the cross-correlation tensor ${\cal R}_{\rm xz}$. From \eqref{eq4}, the $(m,n,p,q)$-th element of ${\cal R}_{\rm xz}$ can be found in the matrix ${\bf X} \otimes {\bf Z}^*$, and it is given by
\begin{equation} \label{eq59}
	\begin{aligned}
		\left[ {\cal R}_{\rm xz} \right]_{mnpq}  & \triangleq x_{mn}z_{pq}^* \\ 
		& =  {\tilde x}_{mn}{\tilde z}^*_{pq} + {w}_{mn}w^*_{pq} + {\tilde x}_{mn}w^*_{pq} + {w}_{mn}{\tilde z}^*_{pq},
	\end{aligned}
\end{equation}
where the first two components ${\tilde x}_{mn}{\tilde z}^*_{pq}$ and ${w}_{mn}w^*_{pq}$ correspond to the received signal of multiple sources and noise, respectively, while the sum of the last two components ${\tilde x}_{mn}w^*_{pq} + {w}_{mn}{\tilde z}^*_{pq}$ represent the {\it cross term} that is generated by the cross-correlation between the signal and noise components. 

Taking first the signal components $\{ {\tilde x}_{mn}{\tilde z}^*_{pq} \}_{m,n,p,q}$ in \eqref{eq59}, we can see that they all can be found in the following matrix
\begin{equation} \label{eq60}
	\begin{aligned}
		{\bf \tilde X} \otimes {\bf \tilde Z}^* & = \left({\dot {\bf A}}_{\rm x}{\bf R}_{\rm s} {\bf K}_{\rm x}^T\right) \otimes \left({\dot {\bf A}}_{\rm z}{\bf R}_{\rm s} {\bf K}_{\rm z}^T\right)^*\\
		& = \left({\dot {\bf A}_{\rm x}} \otimes {\dot {\bf A}_{\rm z}^*}\right)\left({\bf R}_{\rm s} \otimes {\bf R}_{\rm s}\right)\left({{\bf K}_{\rm x}} \otimes {{\bf K}_{\rm z}^*}\right)^T\\
		& = \left({\bf \bar A}_{\rm x} \odot {\bf \bar A}^*_{\rm z}\right){\rm diag}({\bf \bar p})\left({\bf \bar K}_{\rm x} \odot {\bf \bar K}^{*}_{\rm z}\right)^T,
	\end{aligned}
\end{equation}
where ${\bf \bar p} \triangleq {\rm vec}\{{\bf p}{\bf p}^T\}$, ${\bar {\bf A}}_{\rm x} \triangleq {\dot {\bf A}}_{\rm x} \otimes{\bf 1}_{1 \times K} \in {\mathbb{C}^{Q \times K^2}}$, ${\bar {\bf K}}_{\rm x} \triangleq {{\bf K}}_{\rm x} \otimes {\bf 1}_{1 \times K} \in {\mathbb{C}^{M \times K^2}}$, ${\bar {\bf A}}_{\rm z} \triangleq {\bf 1}_{1 \times K} \otimes {\dot {\bf A}}_{\rm z} \in {\mathbb{C}^{Q \times K^2}}$, and ${\bar {\bf K}}_{\rm z} \triangleq {\bf 1}_{1 \times K} \otimes {{\bf K}}_{\rm z} \in {\mathbb{C}^{M \times K^2}}$. Note that the property $\left( {{\bf{AB}}} \right) \otimes \left( {{\bf{CD}}} \right) = \left( {{\bf{A}} \otimes {\bf{C}}} \right)\left( {{\bf{B}} \otimes {\bf{D}}} \right)$ has been used in \eqref{eq60}. Moreover, it can be found that \eqref{eq60} corresponds to the matricized version of the 4-th order tensor, which fits the tensor CPD model \cite{28,30}, given by
\begin{equation}\label{eq61}
	\begin{aligned}
		& {\cal \tilde R}_{\rm xz} = \left[\left[{\bf \bar p}; {\bar {\bf A}}_{\rm x}, {\bar {\bf K}}_{\rm x}, {\bar {\bf A}}_{\rm z}^*, {\bar {\bf K}}_{\rm z}^*\right]\right]\\
		& \left[{\cal \tilde R}_{\rm xz}\right]_{mnpq} \!\!=\!\! {{\sum_{k_1 = k_2 = k} \!\! p_k^2 e^{-j\pi (m+n-S)\cos\theta_k}e^{j\pi (p+q-S)\cos\phi_k}}} \\
		& + \sum_{k_1 \neq k_2} {p_{k_1}p_{k_2}} e^{-j\pi(m+n-S)\cos\theta_{k_1}}e^{j\pi(p+q-S)\cos\phi_{k_2}}.\\
	\end{aligned}
\end{equation}
Then the correspondence between the positions of the elements in ${\bf \tilde X} \otimes {\bf \tilde Z}^*$ and ${\cal \tilde R}_{\rm xz}$, which both consist of the same elements, can be described in terms of the following equality
\begin{equation}\label{eq62}
	{\bf \tilde X} \otimes {\bf \tilde Z}^* = {\rm reshape} \left({\cal \tilde R}_{\rm xz}, \left[\{3,1\}, \{4,2\}\right]\right).
\end{equation}

Due to the inherent structures in the factor matrices of the matrix ${\cal \tilde R}_{\rm xz}$, it can be seen that the designed tensor model \eqref{eq61} has two parts. The first part contains $K$ components that pair the azimuth and elevation angles of sources correctly, while the second part consists of $K(K-1)$ components that mismatch the azimuth and elevation angles of sources. 

For the noise components $\{ w_{mn}w_{pq}^* \}_{m,n,p,q}$, the derivations are similar. Assuming $M > S$, the rank of ${\bf W}$ is $Q$, and we can write that
\begin{equation}\label{eq63}
	\begin{aligned}
		{\bf W} \otimes {\bf W}^* &= \left({\bf I}_Q \otimes {\bf 	I}_Q\right)\left({\bf I}_Q \otimes {\bf I}_Q\right) \left({\bf W}^T \otimes {\bf W}^T\right)^T\\
		& = \left({\bf R}_{ \rm n}^{(1)} \odot {\bf R}_{ \rm n}^{(3)}\right){\rm diag}({\bf 1}_{Q^2 \times 1}) \left({\bf R}_{ \rm n}^{(2)} \odot {\bf R}_{ \rm n}^{(4)}\right)^T,
	\end{aligned}
\end{equation}
where ${\bf R}_{ \rm n}^{(1)} \triangleq {\bf I}_Q \otimes {\bf 1}_{1 \times Q}$, ${\bf R}_{ \rm n}^{(2)} \triangleq {\bf W}^T \otimes {\bf 1}_{1 \times Q}$, ${\bf R}_{ \rm n}^{(3)} \triangleq {\bf 1}_{1 \times Q} \otimes {\bf I}_Q$, and ${\bf R}_{ \rm n}^{(4)} \triangleq {\bf 1}_{1 \times Q} \otimes {\bf W}^T$. Hence, the corresponding noise tensor ${\cal R}_{\rm n}$ satisfies
\begin{equation} \label{eq64}
	\begin{aligned}
		&{\bf W} \otimes {\bf W}^* = {\rm reshape} \left({\cal R}_{\rm n}, \left[\{3,1\}, \{4,2\}\right]\right)\\
		&{\cal R}_{\rm n} = \left[\left[{\bf 1}_{Q^2 \times 1}; {\bf R}_{ \rm n}^{(1)}, {\bf R}_{ \rm n}^{(2)}, {\bf R}_{ \rm n}^{(3)}, {\bf R}_{ \rm n}^{(4)}\right]\right].\\
	\end{aligned}
\end{equation}

Finally, for the cross term components in \eqref{eq59}, we similarly have
\begin{equation}\label{eq65}
	\begin{aligned}
		{\bf \tilde X} \otimes {\bf W}^* & = \left({\dot {\bf A}_{\rm x}} \otimes {\bf I}_Q \right)\left({\bf R}_{\rm s} \otimes {\bf I}_{Q}\right)\left({{\bf K}_{\rm x}} \otimes {{\bf W}^T}\right)^T\\
		& = \left({\bf R}_{ \rm ct,1}^{(1)} \odot {\bf R}_{ \rm ct, 1}^{(3)}\right){\rm diag}({\bf p}_{\rm ct,1}) \left({\bf R}_{ \rm ct, 1}^{(2)} \odot {\bf R}_{ \rm ct,1}^{(4)}\right)^T\\
		{\bf W} \otimes {\bf \tilde Z}^* & = \left({\bf I}_Q \otimes {\dot {\bf A}_{\rm z}^*} \right)\left({\bf I}_{Q} \otimes {\bf R}_{\rm s} \right)\left({{\bf W}^T} \otimes {{\bf K}_{\rm z}^*} \right)^T\\
		& = \left({\bf R}_{ \rm ct,2}^{(1)} \odot {\bf R}_{ \rm ct, 2}^{(3)}\right){\rm diag}({\bf p}_{\rm ct,2}) \left({\bf R}_{ \rm ct, 2}^{(2)} \odot {\bf R}_{ \rm ct,2}^{(4)}\right)^T,\\
	\end{aligned}
\end{equation}
where ${\bf R}_{ \rm ct,1}^{(1)} \triangleq {\dot {\bf A}_{\rm x}} \otimes {\bf 1}_{1 \times Q}$, ${\bf R}_{ \rm ct,1}^{(2)} \triangleq {{\bf K}_{\rm x}} \otimes {\bf 1}_{1 \times Q}$, ${\bf R}_{ \rm ct,1}^{(3)} \triangleq {\bf 1}_{1 \times K} \otimes {{\bf I}_{Q}}$, ${\bf R}_{ \rm ct,1}^{(4)} \triangleq {\bf 1}_{1 \times K} \otimes {{\bf W}^T}$, ${\bf R}_{ \rm ct,2}^{(1)} \triangleq {{\bf I}_{Q}} \otimes {\bf 1}_{1 \times K}$, ${\bf R}_{ \rm ct,2}^{(2)} \triangleq {{\bf W}^T} \otimes {\bf 1}_{1 \times K}$, ${\bf R}_{ \rm ct,2}^{(3)} \triangleq {\bf 1}_{1 \times Q} \otimes {\dot {\bf A}_{\rm z}^*}$, ${\bf R}_{ \rm ct,2}^{(4)} \triangleq {\bf 1}_{1 \times Q} \otimes {{\bf K}_{\rm z}^*}$, ${\bf p}_{\rm ct,1} \triangleq {\bf p} \otimes {\bf 1}_{Q \times 1}$, and ${\bf p}_{\rm ct,2} \triangleq {\bf 1}_{Q \times 1} \otimes {\bf p}$.
Similar to \eqref{eq62} and \eqref{eq64}, we have
\begin{equation}\label{eq66}
	\begin{aligned}
		& {\bf \tilde X} \otimes {\bf W}^* = {\rm reshape}\left({\cal R}_{\rm ct, 1}, \left[\{3,1\}, \{4,2\}\right]\right)\\
		&  {\bf W} \otimes {\bf \tilde Z}^* = {\rm reshape}\left({\cal R}_{\rm ct,2}, \left[\{3,1\}, \{4,2\}\right]\right)\\
		& {\cal R}_{\rm ct,1} = \left[\left[{\bf p}_{\rm ct,1}; {\bf R}_{ \rm ct,1}^{(1)}, {\bf R}_{ \rm ct,1}^{(2)}, {\bf R}_{ \rm ct,1}^{(3)}, {\bf R}_{ \rm ct,1}^{(4)}\right]\right]\\
		& {\cal R}_{\rm ct ,2} = \left[\left[{\bf p}_{\rm ct,2}; {\bf R}_{ \rm ct,2}^{(1)}, {\bf R}_{ \rm ct,2}^{(2)}, {\bf R}_{ \rm ct,2}^{(3)}, {\bf R}_{ \rm ct,2}^{(4)}\right]\right].\\
	\end{aligned}
\end{equation}

Using \eqref{eq60}-\eqref{eq66}, the relationship between ${\cal R}_{\rm xz}$ and ${\bf X} \otimes {\bf Z}^*$ can be given by
\begin{equation} \label{eq14}
	\begin{aligned}
		& {\bf X} \otimes {\bf Z}^* = {\rm reshape}\left({\cal R}_{\rm xz}, \left[\{3,1\}, \{4,2\}\right]\right)\\
		& {\cal R}_{\rm xz} = {\cal \tilde R}_{\rm xz} + {\cal R}_{\rm ct,1} + {\cal R}_{\rm ct,2} + {\cal R}_{\rm n}.
	\end{aligned}
\end{equation}

Consequently, a 4-th order tensor ${\cal R}_{\rm xz}$ can be built from the Kronecker product of matrices ${\bf X}$ and ${\bf Z}^*$ by adjusting the order of all of its elements. In \eqref{eq14}, ${\cal \tilde R}_{\rm xz}$ and ${\cal R}_{\rm n}$ correspond to the received signal of multiple sources and noise, respectively. The last two terms ${\cal R}_{\rm ct,1}$ and ${\cal R}_{\rm ct,2}$ represent the undesirable cross products between the correlated signal and noise components, which together are named as {\it cross term}.
As a consequence, ${\cal R}_{\rm xz}$ can be constructed by changing the order of the elements in ${\bf X} \otimes {\bf Z}^*$. Accordingly, the other cross-correlation tensor ${\cal R}_{\rm zx}$ can also be built by adjusting the order of the elements in ${\bf X}^* \otimes {\bf Z}$. The structure of ${\cal R}_{\rm zx}$ can be observed analogously by replacing ${\dot {\bf A}}_{\rm x}, {\bf K}_{\rm x}, {\dot {\bf A}}_{\rm z}^*, {\bf K}_{\rm z}^*$ with ${\dot {\bf A}}_{\rm x}^*, {\bf K}_{\rm x}^*, {\dot {\bf A}}_{\rm z}, {\bf K}_{\rm z}$ in \eqref{eq60}-\eqref{eq66}, respectively.}

Using both cross-correlation tensors ${\cal R}_{\rm xz}$ and ${\cal R}_{\rm zx}$, a higher-order tensor model that fully exploits the multi-dimensional structure shared by all subarrays in {  the} L-shaped nested array as well as the joint sources spatial information can be designed. 

To demonstrate this, let us take the signal component as an example. {   The tensors of ${\cal \tilde R}_{\rm xz}$ and ${\cal \tilde R}_{\rm zx}$ are given by
\begin{equation}\label{10}
	\begin{aligned}
		& {\cal \tilde R}_{\rm xz} = \left[\left[{\bf \bar p}; {\bar {\bf A}}_{\rm x}, {\bar {\bf K}}_{\rm x}, {\bar {\bf A}}_{\rm z}^*, {\bar {\bf K}}_{\rm z}^*\right]\right]\\
		& {\cal \tilde R}_{\rm zx} = \left[\left[{\bf \bar p}; {\bar {\bf A}}_{\rm x}^*, {\bar {\bf K}}_{\rm x}^*, {\bar {\bf A}}_{\rm z}, {\bar {\bf K}}_{\rm z}\right]\right] .
	\end{aligned}
\end{equation}
Since ${\bar {\bf A}}_{\rm x}, {\bar {\bf K}}_{\rm x}, {\bar {\bf A}}_{\rm z}$ and ${\bar {\bf K}}_{\rm z}$ are Vandermonde matrices, the conjugate symmetric property can be utilized.} It is given by
\begin{equation}\label{11}
    \begin{aligned}
        & {\bf J}_Q{\bf \bar A}_{\rm x} = {\bf \bar A}_{\rm x}^*{\bm \Phi}_{\rm x}, \quad {\bf J}_M{\bf \bar K}_{\rm x} = {\bf \bar K}^*_{\rm x} {\bm \Pi}_{\rm x},\\
        & {\bf J}_Q{\bf \bar A}_{\rm z}= {\bf \bar A}_{\rm z}^{*}{\bm \Phi}_{\rm z}, \quad {\bf J}_M {\bf \bar K}_{\rm z} = {\bf \bar K}_{\rm z}^*{\bm \Pi}_{\rm z},
    \end{aligned}
\end{equation}
where ${\bf J}_Q$ and ${\bf J}_M$ are the exchange matrices of size $Q \times Q$ and $M \times M$, respectively. The diagonal matrices ${\bm \Phi}_{\rm x}$, ${\bm \Phi}_{\rm z}$, ${\bm \Pi}_{\rm x}$ and ${\bm \Pi}_{\rm z}$ are given by
\begin{equation} \label{12}
    \begin{aligned}
        & {\bm \Phi}_{\rm x} \triangleq {\rm diag}\left[{\bm \kappa}_{\rm x}^{-M+3} \otimes {\bf 1}_{1 \times K}\right], \; {\bm \Pi}_{\rm x} \triangleq {\rm diag}\left[{\bm \kappa}_{\rm x}^{M-1} \otimes {\bf 1}_{1 \times K}\right],\\
        & {\bm \Phi}_{\rm z} \triangleq {\rm diag}\left[{\bf 1}_{1\times K} \otimes {\bm \kappa}_{\rm z}^{-M+3}\right], \; {\bm \Pi}_{\rm z} \triangleq {\rm diag}\left[{\bf 1}_{1\times K}\otimes{\bm \kappa}_{\rm z}^{M-1}\right],\\
    \end{aligned}
\end{equation}
with ${\bm \kappa}_{\rm z} \triangleq [e^{-j\pi\cos\phi_1}, \cdots, e^{-j\pi\cos\phi_K}]^T \in {\mathbb{C}^K}$.

Inserting \eqref{11} into ${\cal \tilde R}_{\rm zx}$ and concatenating both ${\cal \tilde R}_{\rm xz}$ and ${\cal \tilde R}_{\rm zx}$ in a new dimension\footnote{{  Here, the elements in ${\cal \tilde R}_{\rm zx}$ are reversed alone all dimensions before the concatenation.}}, the following 5-order tensor of size ${Q \times M \times Q \times M \times 2}$ can be constructed
\begin{equation}\label{13}
    {\cal \tilde R} = \left[\left[{\bf \bar p}; {\bf \bar A}_{\rm x}, {\bf \bar K}_{\rm x}, {\bf \bar A}_{\rm z}^{*}, {\bf \bar K}_{\rm z}^*, {\bf G}\right]\right],
\end{equation}
where ${\bf G} \in {\mathbb{C}^{2 \times K^2}}$ represents the joint sources spatial information in the 5-th dimension, given by
\begin{equation}\label{14}
     {\bf G} \triangleq \left[{\bf 1}_{K^2 \times 1},  {\bf g}\right]^T, \quad {\bf g} \triangleq \left({\bm \kappa_{{\rm x}}^2}\right)^* \otimes {\bm \kappa}_{\rm z}^2.
\end{equation}

Thus, by exploiting the conjugate symmetry property, the effective array aperture is increased and the 5-th factor matrix ${\bf G}$ is built.  The factor matrices of ${\cal \tilde R}$ contain the sources' 2-D DOA information in two dimensions jointly or separately, which can be used to conduct 2-D DOA estimation. However, if conventional tensor decomposition methods like CPD or higher-order SVD (HOSVD) are directly utilized to conduct 2-D DOA estimation \cite{28,30}, the computational complexity may be extremely high. To reduce the computational complexity, we use tensor reshape operator to obtain a 3-order tensor.

Note that different reshapes are not equivalent from the parameter identifiability point of view \cite{29}. We reshape ${\cal \tilde R}$ into a new 3-order tensor such that the system DOF is maximized, and denote this reshape as ${\cal \tilde T} = {\rm reshape}({\cal \tilde R},[\{3,1\},\{4,2\},\{5\}]) \in \mathbb{C}^{Q^2 \times M^2 \times 2}$, or equivalently, as
\begin{equation}\label{15}
    {\cal \tilde T} = \left[\left[{\bf \bar p}; ({\bf \bar A}_{\rm x} \odot {\bf \bar A}_{\rm z}^{*}),({\bf \bar K}_{\rm x} \odot {\bf \bar K}_{\rm z}^*), {\bf G}\right]\right],
\end{equation}
where ${\bf \bar A}_{\rm x} \odot {\bf \bar A}_{\rm z}^{*}$, ${\bf \bar K}_{\rm x} \odot {\bf \bar K}_{\rm z}^*$ and ${\bf G}$ are the first, second and third factor matrices of ${\cal \tilde T}$, respectively. In matrix form, the received signal can be expressed by the tensor unfolding, i.e, ${\bf \tilde T}_{(2)} = {\rm unfolding}({\cal \tilde T},[\{3,1\},\{2\}])$, or equivalently, by
\begin{equation} \label{16}
   {\bf \tilde T}_{(2)} = \left[({\bf \bar A}_{\rm x} \odot {\bf \bar A}_{\rm z}^{*}) \odot {\bf G}\right]{\bf \bar R}_{\rm s}\left({\bf \bar K}_{\rm x} \odot {\bf \bar K}_{\rm z}^*\right)^T,
\end{equation}
where ${\bf \bar R}_{\rm s}  = {\rm diag}({\bf \bar p})$. Then the 2-D DOA estimation problem for {  the} L-shaped nested array consists of finding $\{(\theta_k,\phi_k)\}_{k=1}^K$ from the observation of ${\cal T} \in \mathbb{C}^{Q^2 \times M^2 \times 2}$. The structures of the cross term and the noise term will be introduced in the next section.

\subsection{Parameter Identifiability}
{  As it can be seen from \eqref{eq61}, the construction of ${\cal R}_{\rm xz}$ (and similarly ${\cal R}_{\rm zx}$) takes the advantage of the multi-dimensional structure of the received signal for all subarrays at the cost of introducing $K(K-1)$ additional false targets (see the second summation term in \eqref{eq61}) that may mismatch the azimuth and elevation angles of sources. This can also be found in the structures of the factor matrices of ${\cal \tilde T}$.} 

Nevertheless, the parameter identifiability for our designed tensor model is related to the tensor rank, whose upper bound is restricted by the uniqueness condition of tensor decomposition. The DOA estimation based on ${\cal T}$ can also be regarded as a multi-dimensional harmonic retrieval problem. For the latter problem, the parameter identifiability has been deeply studied \cite{45,46}. For tensors with arbitrary factor matrix, conventional alternating least squares (ALS) algorithm can be used and the uniqueness condition is determined by the sum of the Kruskal ranks of all factor matrices \cite{28}. For tensors with structured factor matrix like Vandermonde matrix, a computationally efficient tensor decomposition method and a better uniqueness condition have been discussed \cite{36,41}. In our case, the uniqueness condition can be given by
\begin{equation}\label{17}
    \min \{2(Q^2-1),M^2\} \ge {  K^2}.
\end{equation}

To further explain the essence of \eqref{17} and to demonstrate the distinct tensor reshape \eqref{15}, rewrite ${\bf \tilde T}_{(2)}$ as
\begin{equation}\label{18}
    {\bf \tilde T}_{(2)} = \sum_{k=1}^{K^2} {\bar p}_k\left[\left({\bf \bar a}_{{\rm x},k} \odot {\bf \bar a}_{{\rm z},k}^{*}\right) \odot {\bf g}_k\right]\left({\bf \bar k}_{{\rm x},k} \odot {\bf \bar k}_{{\rm z},k}^*\right)^T.
\end{equation}

The expression $\left({\bf \bar a}_{{\rm x},k} \odot {\bf \bar a}_{{\rm z},k}^{*}\right) \odot {\bf g}_k$ can be regarded as a steering vector, which corresponds to a virtual co-array that consists of two centrally symmetric URAs. The structure of each URA merely depends on the manifold of ${\bf \bar a}_{{\rm x},k}\odot {\bf \bar a}_{{\rm z},k}^{*}$, since ${\bf g}_k$ is generated by exploiting the conjugate symmetry property. {  Consequently, the maximum number of sources that can be resolved by the virtual co-array is ${\sqrt{2(Q^2-1)}}$, if the number of snapshots is large enough. However, the expression ${\bf \bar k}_{{\rm x},k} \odot {\bf \bar k}_{{\rm z},k}^*$ implies that the number of snapshots is $M^2$ and it is comparable with $Q^2$. In this case, the maximum number of sources that can be resolved by the virtual co-array is ${\sqrt{\min \{2(Q^2-1),M^2\}}}$, which is identical to \eqref{17}.}

It is also worth noting that the aperture of the virtual co-array rises with the increase of $Q$ while the number of efficient snapshots declines. It is typically determined in the conventional nested array that the selection of $Q = M =S$ is optimal in terms of the trade-off between robustness and spatial resolution. In our tensor model, however, the following optimization problem is built to maximize the system DOF
\begin{equation}\label{19}
    \begin{aligned}
        & \max_Q \quad {\sqrt{\min \{2(Q^2-1),M^2\}}}\\
        & {\rm s.t.} \quad Q+M = 2S,
    \end{aligned}
\end{equation}
{ whose optimal solution and optimal value are $Q = \sqrt{8S^2+2}-2S$ and ${\sqrt{24S^2-8S\sqrt{8S^2+2}+2}}$, respectively. Using this result, approximately $\sqrt{1.38S^2+2}$ (17~\% improvement) sources can be resolved with only $2N$ physical elements based on the above designed tensor model.} It is superior to the conventional approaches that treat the received signals for different subarrays in co-array domain separately. 

\section{Proposed Iterative 2-D DOA Estimation Method for L-shaped Nested Array}\label{sec4}
In the previous section, a higher-order tensor signal model has been constructed and a special type of tensor reshape has been utilized to improve the parameter identifiability of the constructed tensor model. Note that the first two factor matrices of ${\cal \tilde T}$ are the KR product of two Vandermonde matrices, whose vectors of generators contain the {  sources' angular information}. We can thus exploit the shift-invariance between different subarrays in two axes to conduct the 2-D DOA estimation. However, in the co-array domain, the signal and noise terms become correlated, and the cross term between them cannot be ignored (see, for example, \cite{22}). 

Using the same operations as we did for deriving \eqref{13} and \eqref{15}, the cross term between signal and spatially correlated noise as well as the noise term can be also expressed in the tensor form as ${\cal T}_{\rm ct}$ and ${\cal T}_{\rm n}$. {   The matricized version of the corresponding tensors, which are ${\bf T}_{{\rm ct},(2)} = {\rm unfolding}\left({\cal  T}_{\rm ct},[\{3,1\},\{2\}]\right) \in {\mathbb{C}^{2Q^2 \times M^2}}$ and ${\bf T}_{{\rm n},(2)} = {\rm unfolding}\left({\cal  T}_{\rm n},[\{3,1\},\{2\}]\right) \in {\mathbb{C}^{2Q^2 \times M^2}}$, can be written as
\begin{equation}\label{20}
    {\bf T}_{{\rm ct},(2)} = {\bf P}\left[{\bf D}, \, {\bf D}^*\right]^T, \quad {\bf T}_{{\rm n},(2)} = {\bf P}\left[{\bf N}, \, {\bf N}^*\right]^T,
\end{equation}
where ${\bf D} \triangleq \left({\bf \tilde X} \otimes {\bf W} + {\bf W} \otimes {\bf \tilde Z}\right)^T$, ${\bf N} \triangleq \left({\bf W} \otimes {\bf W}\right)^T$, and ${\bf P} \in {\mathbb{R}^{2Q^2 \times 2Q^2}}$ is a permutation matrix that sequentially takes out odd and even rows of a matrix to build a new matrix}. Hence, the unfolding of the designed tensor that contains the signal term, noise term and cross term components can be written as
\begin{equation}\label{21}
    {\bf T}_{(2)} = {\bf \tilde T}_{(2)} + {\bf T}_{{\rm ct},(2)} + {\bf T}_{{\rm n},(2)}.
\end{equation}

Compared with the conventional technique based on averaging the signal covariance matrices of all subarrays, the spatially correlated cross term ${\bf T}_{{\rm ct},(2)}$ degrades the DOA estimation performance. Note that both ${\bf T}_{{\rm ct},(2)}$ and ${\bf T}_{{\rm n},(2)}$ are sparse matrices, i.e., most of their elements are zeros. To resolve the aforementioned problem of performance degradation caused by the cross term, an iterative DOA estimation method is proposed next. The main idea of the method is to modify the received signal at every next step of the estimation procedure based on the DOA estimation results obtained in the previous step \cite{37}. Thus, the DOA estimation performance can be improved by estimating and removing the cross term in the received signal iteratively.

\subsection{Step~1: DOA Estimation via Tensor Decomposition with Vandermonde Factor Matrix}
Given the received signal matrices ${\bf X}$ and ${\bf Z}$, the 3-order tensor ${\cal T}$ can be constructed, whose factor matrices are Vandermonde matrices in the noise-less case. Since we assume that all sources are spatially distinct, both ${\bf \bar A}_{\rm x} \odot {\bf \bar A}_{\rm z}^{*}$ and ${\bf \bar K}_{\rm x} \odot {\bf \bar K}_{\rm z}^*$ are {   full column rank}. Hence, a computationally efficient tensor decomposition method can be designed \cite{36,41}.

Specifically, consider the matrix in \eqref{16}. Denote the truncated SVD of this matrix as\footnote{The truncated SVD returns the dominant singular vectors and the associated singular values of a matrix.} ${\bf T}_{(2)} = {\bf U}{\bf \Lambda}{\bf V}^H$ , where ${\bf U} \in \mathbb{C}^{2Q^2 \times K^2}$, ${\bf \Lambda} \in \mathbb{C}^{K^2 \times K^2}$, and ${\bf V} \in \mathbb{C}^{M^2 \times K^2}$. Since ${\bf \bar A}_{\rm x} \odot {\bf \bar A}_{\rm z}^{*}$ and ${\bf \bar K}_{\rm x} \odot {\bf \bar K}_{\rm z}^*$ are {   full column rank}, for a nonsingular matrix ${\bf \Xi} \in \mathbb{C}^{K^2 \times K^2}$, it can be found that
\begin{equation}\label{22}
    {\bf U}{\bf \Xi} = \left({\bf \bar A}_{\rm x} \odot {\bf \bar A}_{\rm z}^{*}\right) \odot {\bf G}.
\end{equation}

Considering the KR product, the following relationships hold
\begin{equation}\label{23}
    \begin{aligned}
        & {\bf U}_{{\rm x}1}{\bf \Xi} = ({\bf \bar A}_{{\rm x}1} \odot {\bf \bar A}_{\rm z}^{*}) \odot {\bf G}, \quad {\bf U}_{{\rm x}2}{\bf \Xi} = ({\bf \bar A}_{{\rm x}2} \odot {\bf \bar A}_{\rm z}^{*}) \odot {\bf G},\\
        & {\bf U}_{{\rm z}1}{\bf \Xi} = ({\bf \bar A}_{\rm x} \odot {\bf \bar A}_{{\rm z}1}^{*}) \odot {\bf G}, \quad {\bf U}_{{\rm z}2}{\bf \Xi} = ({\bf \bar A}_{\rm x} \odot {\bf \bar A}_{{\rm z}2}^{*}) \odot {\bf G},\\
    \end{aligned}
\end{equation}
where ${\bf \bar A}_{{\rm x}2}$ and ${\bf \bar A}_{{\rm z}2}$ denote the submatrices of ${\bf \bar A}_{\rm x}$ and ${\bf \bar A}_{\rm z}$ without the first row, respectively, ${\bf \bar A}_{{\rm x}1}$ and ${\bf \bar A}_{{\rm z}1}$ denote the submatrices of ${\bf \bar A}_{\rm x}$ and ${\bf \bar A}_{\rm z}$ without the last row, respectively, ${\bf U}_{{\rm x}1}$, ${\bf U}_{{\rm x}2}$, ${\bf U}_{{\rm z}1}$ and ${\bf U}_{{\rm z}2}$ are the submatrices of the left singular matrix ${\bf U}$, given by
\begin{equation}\label{24}
    \begin{aligned}
        & {\bf U}_{{\rm x}1} \triangleq \left[{\bf I}_{2Q(Q-1)}, {\bf 0}_{2Q(Q-1) \times 2Q}\right]{\bf U} \\
        & {\bf U}_{{\rm x}2} \triangleq \left[{\bf 0}_{2Q(Q-1) \times 2Q}, {\bf I}_{2Q(Q-1)}\right]{\bf U} \\
        & {\bf U}_{{\rm z}1} \triangleq \left({\bf I}_Q \otimes \left[{\bf I}_{2(Q-1)}, {\bf 0}_{2(Q-1) \times 2}\right]\right){\bf U} \\
        & {\bf U}_{{\rm z}2} \triangleq \left({\bf I}_Q \otimes \left[{\bf 0}_{2(Q-1) \times 2}, {\bf I}_{2(Q-1)}\right]\right){\bf U}. \\
    \end{aligned}
\end{equation}

{  Note that ${\bf \bar A}_{{\rm x}2} = {\bf \bar A}_{{\rm x}1}{\bf \Gamma}_{\rm x}$ and ${\bf \bar A}_{{\rm z}2} = {\bf \bar A}_{{\rm z}1}{\bf \Gamma}_{\rm z}$, where ${\bf \Gamma}_{\rm x} \triangleq {\rm diag}\left({\bm \kappa}_{\rm x} \otimes {\bf 1}_{K \times 1}\right)$ and ${\bf \Gamma}_{\rm z} \triangleq {\rm diag}\left({\bf 1}_{K \times 1} \otimes {\bm \kappa}_{\rm z} \right)$. Using these properties in \eqref{23}, we can write that
\begin{equation}\label{25}
    {\bf U}_{{\rm x}2}{\bf \Xi} = {\bf U}_{{\rm x}1}{\bf \Xi}{\bf \Gamma}_{\rm x}, \quad {\bf U}_{{\rm z}2}{\bf \Xi} = {\bf U}_{{\rm z}1}{\bf \Xi}{\bf \Gamma}_{\rm z}^{*},
\end{equation}
or equivalently,
\begin{equation}\label{26}
    {\bf U}_{{\rm x}1}^{\dag}{\bf U}_{{\rm x}2} = {\bf \Xi}{\bf \Gamma}_{\rm x}{\bf \Xi}^{-1}, \quad {\bf U}_{{\rm z}1}^{\dag}{\bf U}_{{\rm z}2} = {\bf \Xi}{\bf \Gamma}_{\rm z}^{*}{\bf \Xi}^{-1}.
\end{equation}

From \eqref{26}, the eigenvalues of the matrices ${\bf U}_{{\rm x}1}^{\dag}{\bf U}_{{\rm x}2}$ and ${\bf U}_{{\rm z}1}^{\dag}{\bf U}_{{\rm z}2}$ can be regarded as the estimations of the diagonal elements of ${\bf \Gamma}_{\rm x}$ and ${\bf \Gamma}_{\rm z}^{*}$, respectively. Taking the $K^2$ eigenvalues of ${\bf U}_{{\rm x}1}^{\dag}{\bf U}_{{\rm x}2}$, for example, there are only $K$ unique eigenvalues while the other $K(K-1)$ eigenvalues are repeated. Thus, the unique eigenvalues still need to found, and they can be found by the k-means clustering algorithm \cite{50, 51}, for example. Such unique eigenvalues can be regarded as the estimates of the elements of ${\bm \kappa}_{\rm x}$. The estimation of ${\bm \kappa}_{\rm z}$ is similar.} After estimating ${\bm \kappa}_{\rm x}$ and ${\bm \kappa}_{\rm z}$, the angles ${\hat \theta_k}$ and ${\hat \phi_k} $ can be computed by
\begin{equation}\label{27}
    {\hat \theta_k} = \arccos\left({j\ln{{\hat \kappa}_{{\rm x},k}/\pi}}\right), \quad {\hat \phi_k} = \arccos\left({j\ln{{\hat \kappa}_{{\rm z},k}/\pi}}\right),
\end{equation}
where ${\hat \kappa}_{{\rm x},k}$ and ${\hat \kappa}_{{\rm z},k}$ are the $k$-th elements of the corresponding vectors. { Although the redundancies in the $K^2$ eigenvalues of ${\bf U}_{{\rm x}1}^{\dag}{\bf U}_{{\rm x}2}$ and ${\bf U}_{{\rm z}1}^{\dag}{\bf U}_{{\rm z}2}$ can be mitigated via k-means clustering algorithm, the pair-matching of the remaining eigenvalues (or equivalently, the estimated elevation and azimuth angles in \eqref{27}) is still required to fully remove the additional $K(K-1)$ false sources that are introduced by the designed tensor model.

Consider the CCM of the received signals in two axes, given by ${\bf R}_{\rm c} \triangleq {\rm E}\{{\bf x}(t){\bf z}^H(t)\} = {\bf A}_{\rm x}{\bf R}_{\rm s}{\bf A}_{\rm z}^H$. Assuming that the order of $\{\hat \phi_k\}_{k = 1}^K$ in \eqref{27} is correct, there exists a permutation matrix ${\bf E} \in {\mathbb{R}^{K \times K}}$ that satisfies 
\begin{equation}\label{eq33}
	{\bf R}_{\rm c} = {\bf A}_{\rm x}{\bf R}_{\rm s}{\bf A}_{\rm z}^H = {\bf \hat A}_{\rm x}{\bf E}{\bf \hat R}_{\rm s}{\bf \hat A}_{\rm z}^H,
\end{equation}
where ${\bf \hat A}_{\rm x}$ and ${\bf \hat A}_{\rm z}$ are constructed using \eqref{27}, $e_{mn} \in \{0,1\}$, $\sum_{m = 1}^{K}e_{mn} = 1$, ${\sum_{n = 1}^{K}}e_{mn} = 1$, and ${\bf \hat R}_{\rm s} = {\rm diag}({\bf \hat p})$. The vector ${\bf p}$ can be estimated by solving the following least-squares (LS) problem
\begin{equation} \label{eq34}
	{\bf \hat p} = \arg \min_{\bf p}||{\bf \ddot y}_{\rm z} - {\bf \ddot A}_{\rm z}{\bf p}||^2.
\end{equation}
The solution to \eqref{eq34} is ${\bf \hat p} = \left({\bf \ddot A}_{\rm z}^H{\bf \ddot A}_{\rm z}\right)^{-1}{\bf \ddot A}_{\rm z}{\bf \ddot y}_{\rm z}$. Then, the estimation of ${\bf E}$ can be found by solving the following LS problem
\begin{equation}\label{eq35}
	{\bf \hat E} = \arg\min_{\bf E} \left|\left|{\bf R}_{\rm c} - {\bf \hat A}_{\rm x}{\bf E}{\bf \hat R}_{\rm s}{\bf \hat A}^H_{\rm z}\right|\right|_{\rm F}.
\end{equation}

Hence, the sources azimuth and elevation angles can be paired by sorting $\{\hat \theta_k\}_{k = 1}^K$ via ${\bf \hat E}$.}

\begin{algorithm}[tb]\label{algorithm}
	\caption{Proposed Iterative 2-D DOA Estimation Method for {  the} L-shaped Nested Array}
	\KwIn{$K$ and observations of ${\bf x}(t)$ and ${\bf z}(t)$}
	\KwOut{$\left \{(\theta_k,\phi_k)\right\}_{k=1}^{k=K}$}
	${\bf R}_{\rm x}, {\bf R}_{\rm z} \leftarrow {\rm E}\{{\bf x}(t){\bf x}^H(t)\}, {\rm E}\{{\bf z}(t){\bf z}^H(t)\}$\;
	${\ddot{\bf y}}_{\rm x}, {\ddot{\bf y}}_{\rm z} \leftarrow$ \eqref{2} and \eqref{3}\;
	${\bf X}, {\bf Z}  \leftarrow [{\ddot{\bf y}}_{\rm x}^{(1)},{\ddot{\bf y}}_{\rm x}^{(2)},\cdots,{\ddot{\bf y}}_{\rm x}^{(M)}] , [{\ddot{\bf y}}_{\rm z}^{(1)},{\ddot{\bf y}}_{\rm z}^{(2)},\cdots,{\ddot{\bf y}}_{\rm z}^{(M)}]$\;
	${\cal R}_{\rm xz}, {\cal R}_{\rm zx} \leftarrow$\eqref{eq14}\;
	{${\cal R} \leftarrow $ concatenate ${\cal R}_{\rm xz}, {\cal R}_{\rm zx}$ in the fifth dimension}\;
	${\cal T} \leftarrow {\rm reshape}({\cal R},[\{3,1\},\{4,2\},\{5\}])$\;
	${\bf T}_{(2)} \leftarrow {\rm unfolding}({\cal T},[\{3,1\},\{2\}])$\;
	$\epsilon^{<0>} = \left|\left|{\bf T}_{(2)} \right|\right|_{\rm F}^2, \delta = 10^{-5}$\;
	\While{${  \left|\epsilon^{<\ell>} - \epsilon^{<\ell-1>} \right| \geq \delta}$ {\rm and} $ {  {\ell}} \le L$}{
		{\textbf{Step~1}} \Begin{
			$\left( {\bf U}, {\bf \Lambda}, {\bf V} \right) \leftarrow \rm{SVD}({\bf T}^{{{ <\ell-1>}}}_{(2)})$\;
			${\bf U}_{{\rm x}1}^{{{  <\ell>}}},{\bf U}_{{\rm x}2}^{{{  <\ell>}}},{\bf U}_{{\rm z}1}^{{{  <\ell>}}},{\bf U}_{{\rm z}2}^{{{  <\ell>}}} \leftarrow$\eqref{24}\;
			${\bm \kappa}_{\rm x}^{{{  <\ell>}}}, {\bm \kappa}_{\rm z}^{{{  <\ell>}}}$ $\leftarrow$ \eqref{26} and clustering algorithm\;
			${\hat \theta}_k^{{{  <\ell>}}}, {\hat \phi}_k^{{{  <\ell>}}} \leftarrow$ \eqref{27}\;
			${\bf \hat p}^{{ <\ell>}} \leftarrow $ \eqref{eq34}\;
			${\bf \hat E}^{{ <\ell>}} \leftarrow$ \eqref{eq35}\;
			$\left \{(\theta_k,\phi_k)\right\}_{k=1}^{k=K} \leftarrow $ pair-matching via ${\bf \hat E}$ \;
		}
		{\textbf{Step~2}} \Begin{
			${\bf \dot A}_{\rm x}^{{{  <\ell>}}},{\bf \dot A}_{\rm z}^{{{  <\ell>}}}, {\bf K}_{\rm x}^{{{  <\ell>}}},{\bf K}_{\rm x}^{{{  <\ell>}}} \leftarrow$ \eqref{33} and \eqref{37}\;
			${\bf \tilde X}^{{{  <\ell>}}}, {\bf \tilde Z}^{{{  <\ell>}}} \leftarrow$ \eqref{36}\;
			${\bf \hat T}^{{{  <\ell>}}}_{{\rm ct},(2)} \leftarrow$ \eqref{38}\;
			${\bf T}^{{{  <\ell>}}}_{(2)} \leftarrow$ \eqref{39}\;
			$\epsilon^{{  <\ell>}} \leftarrow \left|\left| {\bf T}_{(2)}-{\bf T}_{(2)}^{{{  <\ell>}}} \right|\right|_{\rm F}^2$\;
			${  \ell \leftarrow \ell+1}$\;
		}
	}
\end{algorithm}

\subsection{Step~2: Cross Term Estimation and Elimination}
In Step~2, the DOA estimation results from the previous step can be used to build a scaled version of the undesirable cross term ${\hat{\bf T}}_{{\rm ct},(2)}$. The renewed tensor ${\cal T}^{{  <\ell>}}$ after removing the estimated ${\hat{\bf T}}_{{\rm ct},(2)}$ can be then used as an input for Step~1 again to obtain the DOA estimates with a lower error. First, let us build the steering matrices of two difference co-arrays, i.e.,
\begin{equation}\label{33}
    \begin{aligned}
        & {\bf \dot A}_{\rm x}^{{{  <\ell>}}} = \left[{\bf \dot a}_{\rm x}({\hat \theta_{1}}),{\bf \dot a}_{\rm x} ({\hat \theta_{2}}), \cdots, {\bf \dot a}_{\rm x} ({\hat \theta_{K}})\right]\\
        & {\bf \dot A}_{\rm z}^{{{  <\ell>}}} = \left[{\bf \dot a}_{\rm z}({\hat \phi_{1}}),{\bf \dot a}_{\rm z}({\hat \phi_{2}}), \cdots, {\bf \dot a}_{\rm z} ({\hat \phi_{K}})\right],
    \end{aligned}
\end{equation}
where the superscript {  $(\cdot)^{<\ell>}$} stands for the current iteration, {  $\{\hat \theta_k\}_{k = 1}^K$ and $\{\hat \phi_k\}_{k = 1}^K$ are the estimated source azimuth and elevation angles after pair-matching in Step~1.} In the following, we drop the superscript {  $(\cdot)^{<\ell>}$} in our derivations for notation simplicity. The received signal of all subarrays in two axes can be estimated as
\begin{equation}\label{36}
    {\bf \tilde X} = {\bf \dot A}_{\rm x}{\bf \hat R}_{\rm s} {\bf K}_{\rm x}^{T}, \quad {\bf \tilde Z} = {\bf \dot A}_{\rm z}{\bf \hat R}_{\rm s} {\bf K}_{\rm z}^{T},
\end{equation}
where the reconstructed matrices ${\bf K}_{\rm x} \in {\mathbb{C}^{M \times K}}$ and ${\bf K}_{\rm z} \in {\mathbb{C}^{M \times K}}$ for Step~2 are given by
\begin{equation}\label{37}
    \begin{aligned}
        & {\bf K}_{\rm x} \triangleq \left[{\hat {\bm \kappa}}_{\rm x}^{0},{\hat {\bm \kappa}}_{\rm x}^1, \cdots, {\hat {\bm \kappa}}_{\rm x}^{M-1}\right]^T\\
        & {\bf K}_{\rm z} \triangleq \left[{\hat {\bm \kappa}}_{\rm z}^{0},{\hat {\bm \kappa}}_{\rm z}^1, \cdots, {\hat {\bm \kappa}}_{\rm z}^{M-1}\right]^T,
    \end{aligned}
\end{equation}
with ${\hat {\bm \kappa}}_{\rm x} \triangleq \left[e^{-j\pi\cos\hat \theta_{1}}, \cdots, e^{-j\pi\cos\hat \theta_{K}}\right]^T \in {\mathbb{C}^K}$ and ${\hat {\bm \kappa}}_{\rm z} \triangleq \left[e^{-j\pi\cos\hat \phi_{1}},\cdots,e^{-j\pi\cos \hat \phi_{K}}\right]^T \in {\mathbb{C}^K}$. Therefore, the cross term between the signal and the spatially correlated noise is obtained to be
\begin{equation}\label{38}
    {\bf \hat T}_{{\rm ct},(2)} = {\bf P}\left[{\bf \hat D}, {\bf \hat D}^*\right]^T,
\end{equation}
where ${\bf \hat D} \triangleq \left({\bf \tilde X} \otimes {\bf W} + {\bf W} \otimes {\bf \tilde Z}\right)^T$.

Inserting \eqref{38} to \eqref{21}, the updated received signal in Step~2 in the matrix form is given by
\begin{equation}\label{39}
    {\bf T}_{(2)} \leftarrow {\bf T}_{(2)} - {\mu}{\bf \hat T}_{{\rm ct},(2)},
\end{equation}
where $\mu$ is a real number between zero and one, that is, a reliability factor to the estimates in Step~1. Once $\mu$ is determined, the modified received signal ${\bf T}_{(2)}$ can be updated and the DOA estimation with smaller error can be conducted via the tensor decomposition approach used in Step~1. These two steps can be repeated consequently several times until the convergence or until the desired estimation error is achieved.

The scaling factor $\mu$ represents the reliability of the estimates ${\bf \hat T}_{{\rm ct},(2)}$, i.e., if $\mu$ takes a value close to one, we believe that the estimation error of the cross term is negligible, while a small value of $\mu$ implies that the estimates are erroneous. If $\mu = 1$, it means that the cross term can be precisely estimated and removed. In practice, however, estimation errors are unavoidable. One can find a practical method based on the maximum likelihood (ML) criterion to determine the optimal value of $\mu$ \cite{37}.

An outline of the proposed iterative DOA estimation method for {   the} L-shaped nested array is summarized in \textbf{Algorithm~\ref{algorithm}}.

\subsection{Computational Complexity}
We analyze here the computational complexity of the proposed iterative 2-D DOA estimation method. The initial inputs of the proposed method are the received signals from two axes. The complexity of obtaining the designed tensor $\cal T$ and its matrix unfolding ${\bf T}_{(2)}$ is ${\cal O}\{2N^2T_{\rm s}+2Q^2M^2\}$. In Step~1, the proposed method mainly contains three parts, i.e., the truncated SVD of ${\bf T}_{(2)}$, the EVD of two matrices and the pair-matching procedure. If \eqref{19} is satisfied, then $Q < M$ and ${\bf T}_{(2)} \in {\mathbb{C}}^{2Q^2 \times M^2}$ is a tall matrix, and the complexity of SVD in this case is ${\cal O}\{2Q^2M^4\}$. While computing ${\bm \kappa}_{\rm x}$ and ${\bm \kappa}_{\rm z}$, { the number of flops required is ${\cal O} \{16Q^3(Q-1)K^2 + 4Q(Q-1)K^4 + 2K^6\}$. The pair-matching requires ${\cal O} \{2K^2(S-1)+2KN^2\}$ flops. In Step~2, the construction of the estimated cross term ${\bf \hat T}_{{\rm ct},(2)} $ needs ${\cal O} \{2Q^2M^2 + 2QK(M+K)\}$ flops. Then, the computations in Step~1 are conducted again with the updated inputs. For simplicity, let us consider only one iteration of cross term mitigation, i.e., $L = 1$. Then, the number of flops required is approximately ${\cal O} \{2N^2T_{\rm s} + 4Q^2M^2 + 2QK(M+K) + 8Q(Q-1)K^2(4Q^2+K^2) + 4K(K^5+(S-1)K+N^2)\}$}.  {   The construction of ${\bf T}_{(2)}$ and the EVD of two matrices consume the most of the computational complexity. As a consequence, the computational complexity of the proposed method is ${\cal O} \{2N^2T_{\rm s} + 32Q^3(Q-1)K^2\}$.}

In Table~\ref{table1}, we compare the computational complexity of the proposed approach with that of several other DOA estimation methods.
\begin{table}[t]
	{  
	\caption{The computational complexities of different DOA estimation methods for the L-shaped nested array}\label{table1}
	\begin{threeparttable}
	\centering
	\begin{tabular}{ll}
		\toprule
		Methods   & Computational Complexity \\ \midrule
		JSVD \cite{12}    & ${\cal O} \{2N^2T_{\rm s} + 4S(2S-1)^2 +4N_{\rm s} S^2K\}$   \\
		CESA \cite{15}    & ${\cal O} \{2N^2T_{\rm s} +4SK^2(4S-K)+4N_{\rm s} (4S-K)^2\}$\\
		CSAP \cite{18}    & ${\cal O} \{2N^2T_{\rm s} + 128S^3+16N_{\rm s}S^2(4S-K)\}$ \\
		TALA \cite{19}    & ${\cal O} \{2N^2T_{\rm s} +32LS^4K\}$  \\
		SS \cite{22}      &  ${\cal O} \{2N^2T_{\rm s} + 2N_{\rm s} S^2K\}$\\
		CPD \cite{28}     & ${\cal O} \{2N^2T_{\rm s} + 40LS^4K\}$  \\ 
		Proposed & ${\cal O} \{2N^2T_{\rm s} + 16Q^3(Q-1)(L+1)K^2\}$ \\ \bottomrule                 
	\end{tabular}
    \begin{tablenotes}   
		\footnotesize     
		\item[*] $N_{\rm s}$ denotes the total number of searches, while $L$ is the number of iterations.
	\end{tablenotes}        
\end{threeparttable}}
\end{table}

\subsection{CRB for the Proposed Tensor Model}
It is also worth deriving analytical expression for the CRB for the proposed tensor model to see that the improvement comes from the proposed received signal model and also check whether the proposed algorithm achieves the statistical bound. Since CRB is an asymptotic bound, but the cross term is the result of mismatches that have non-asymptotic nature, we just assume for the CRB derivation that the cross term is fully mitigated. It guarantees that we derive the lowest bound for the best achievable performance independent of whether the cross term is present or fully mitigated.

The tensor model \eqref{13} is used to conduct the 2-D DOA estimation. The received signal spatial covariance matrix without the cross term can be written as
\begin{equation}\label{40}
	{\bf R} = \sum_{k=1}^{\bar K} {{\bar p}_k^2}{\bf c}_{\rm x} (\bar \theta_k) {\bf c}_{\rm z}^H(\bar \phi_k) + {\bf R}_{\rm n},
\end{equation}
where ${\bf c}_{\rm x} (\bar \theta_k) \triangleq {\bf \bar a}_{\rm x} (\bar \theta_k) \otimes {\bf \bar k}_{\rm x} (\bar \theta_k) \otimes {\bf g}_{\rm x} (\bar \theta_k)$, ${\bf c}_{\rm z} (\bar \phi_k) \triangleq {\bf \bar a}_{\rm z} (\bar \phi_k) \otimes {\bf \bar k}_{\rm z} (\bar \phi_k) \otimes {\bf g}_{\rm z} (\bar \phi_k)$, ${\bf g}_{\rm x} (\bar \theta_k) \triangleq [1, e^{j2\pi\cos{\bar \theta}_k}]^T$, ${\bf g}_{\rm z} (\bar \phi_k) \triangleq [1, e^{j2\pi\cos{\bar \phi}_k}]^T$, {   ${\bar {\bm \theta}} \triangleq {\left({\bm \theta} \otimes {\bf 1}_{K \times 1}\right)}$, ${\bar {\bm \phi}} \triangleq {\left({\bf 1}_{K \times 1} \otimes {\bm \phi} \right)}$, ${\bm \theta} \triangleq \left[\theta_1, \cdots, \theta_K\right]^T$, ${\bm \phi} \triangleq \left[\phi_1, \cdots, \phi_K\right]^T$, ${\bar K} = K^2$, and ${\bf R}_{\rm n}$ is the noise spatial covariance matrix, denoted by
\begin{equation}
	{\bf R}_{\rm n} \triangleq \left[ \begin{array}{l}
		{{\rm vec}\{{\bf W}\}}\\
		{{\rm vec}\{{\bf W}\}}
	\end{array} \right]\left[ {{{({\rm vec}\{{\bf W}\})^T}},{{({\rm vec}\{{\bf W}\})^T}}} \right] .
\end{equation}
} Vectorizing \eqref{40}, the received signal vector can be written as
\begin{equation}\label{41}
	{\bf r} = {\sum_{k=1}^{\bar K} {{\bar p}_k^2} {\bf c}_{\rm x} (\bar \theta_k) \otimes {\bf c}_{\rm z}^*(\bar \phi_k)} + {\rm vec}\{{\bf R}_{\rm n}\}.
\end{equation}

Let us collect all unknown but deterministic entities to a $3K^2 \times 1$ vector
\begin{equation}\label{42}
	{  {\bm \psi} \triangleq \left[{\bar {\bm \theta}}^T, {\bar {\bm \phi}}^T, {\bf \bar p}^T\right]^T.}
\end{equation}

Using the Slepian-Bangs (SB) formula \cite{39}, the Fisher information matrix (FIM) can be found as
\begin{equation}\label{43}
	\begin{aligned}
		{\bf J}(\bm \psi) & = {T_{\rm s}}{\rm tr}\left( {{\bf{R}}^{ - 1}\frac{{\partial {{\bf{R}}}}}{{\partial {{{\bm{\psi }} }}}}{\bf{R}}^{ - 1}\frac{{\partial {{\bf{R}}}}}{{\partial {{{\bm{\psi }}}}}}} \right) \\
		& = {T_{\rm s}}{\left( {\frac{{\partial {{\bf{r}}}}}{{\partial{\bm{\psi}}}}} \right)^H}\left( {{\bf{R}}^T \otimes {\bf{R}}} \right)^{-1}\left( {\frac{{\partial {{\bf{r}}}}}{{\partial {\bm{\psi }}}}} \right),
	\end{aligned}
\end{equation}
where
\begin{equation}\label{44}
	{  \frac{{\partial {{\bf{r}}}}}{{\partial {\bm{\psi }}}} \triangleq \left[\left({\frac{{\partial {{\bf{r}}}}}{{\partial {{\bar {\bm \theta}}}}}}\right)^T,\left({\frac{{\partial {{\bf{r}}}}}{{\partial {{\bar {\bm \phi}}}}}}\right)^T,\left({\frac{{\partial {{\bf{r}}}}}{{\partial{{\bar {\bf p}}}}}}\right)^T\right]^T.}
\end{equation}

To compute the derivatives \eqref{44}, we only need to consider two submatrices sequentially, i.e.,
\begin{equation}\label{45}
	\frac{{\partial {\bf{r}}}}{{\partial {\bm{\psi }}}} = \left[ {\underbrace {({\bf{C}}_{\rm x}^\prime  \odot {{\bf{C}}_{\rm z}^*}){{\bf{\bar R}}_{\rm s}^2},({\bf{C}}_{\rm z}^{\prime*} \odot {{\bf{C}}_{\rm x}}){{\bf{\bar R}}_{\rm s}^2}}_{{{\bf{C}}^\prime }},\underbrace {{{\bf{C}}_{\rm xz}}{{\bf{\bar R}}_{\rm s}}}_{\bf{C}}} \right],
\end{equation}
where ${\bf C}_{\rm xz} \triangleq 2({\bf C}_{\rm x} \odot {\bf C}_{\rm z}^*)$, ${\bf C}_{\rm x} \triangleq \left[{\bf c}_{\rm x} (\bar \theta_1), \cdots, {\bf c}_{\rm x} (\bar \theta_{\bar K})\right]$, ${\bf C}_{\rm z} \triangleq \left[{\bf c}_{\rm z} (\bar \phi_1), \cdots, {\bf c}_{\rm z} (\bar \phi_{\bar K})\right]$, and
\begin{equation}\label{46}
	\begin{aligned}
		& {\bf{C}}_{\rm x}^\prime \triangleq \left[ {\frac{{\partial {{\bf{c}}_{\rm x}}({\bar \theta _1})}}{{\partial {\bar \theta _1}}},\cdots,\frac{{\partial {{\bf{c}}_{\rm x}}({\bar \theta _{\bar K}})}}{{\partial {\bar \theta_{\bar K}}}}} \right]\\
		& {\bf{C}}_{\rm z}^\prime  \triangleq \left[ {\frac{{\partial {{\bf{c}}_{\rm z}}({\bar \phi _1})}}{{\partial {\bar \phi _1}}},\cdots,\frac{{\partial {{\bf{c}}_{\rm z}}({\bar \phi _{\bar K}})}}{{\partial {\bar \phi _{\bar K}}}}} \right].
	\end{aligned}
\end{equation}

Then, the FIM can be obtained as
\begin{equation}\label{47}
	{\bf{J}}({\bm \psi}) = T_{\rm s} \left[ {\begin{array}{*{20}{c}}
			{{\bf{J}}_1^H{{\bf{J}}_1}}&{{\bf{J}}_1^H{{\bf{J}}_2}}\\
			{{\bf{J}}_2^H{{\bf{J}}_1}}&{{\bf{J}}_2^H{{\bf{J}}_2}}
	\end{array}} \right],
\end{equation}
where ${\bf J}_1 \triangleq \left( {\bf R}^T \otimes {\bf R}\right)^{-1/2}{\bf C}^\prime$ and ${\bf J}_2 \triangleq \left( {\bf R}^T \otimes {\bf R}\right)^{-1/2}{\bf C}$. Considering the inverse of a $2 \times 2$ block matrix, the CRB can be derived as
\begin{equation}\label{48}
	{\rm CRB} ({\bm \psi}) = \frac{1}{{{T_{\rm s}}}}{\left( {{\bf{J}}_1^H{\bf{\Pi }}_{{{\bf{J}}_2}}^{\perp}{{\bf{J}}_1}} \right)^{ - 1}},
\end{equation}
where ${{\bf{\Pi }}_{{{\bf{J}}_2}}^{\perp}} \triangleq {\bf I}_{4Q^2M^2}-{\bf J}_2({\bf J}_2^H{\bf J}_2)^{-1}{\bf J}_2^H$.

\subsection{Discussions on Small Sample Size and Sources Number Estimation}\label{subsection1}
In practice, the sample estimate of the auto-correlation matrix ${\bf R}_{\rm x}$ obtained as ${\bf \hat R}_{\rm x} \approx (1/{T_{\rm s}}) \sum_{t = 1}^{T_{\rm s}} {\bf x}(t){\bf x}^H(t)$ is typically used. When the number of snapshots is large enough, ${\bf \hat R}_{\rm x} \approx {\bf A}_{\rm x} {\bf R}_{\rm s} {\bf A}_{\rm x}^H + \sigma_{\rm n}^2{\bf I}_N$ and \eqref{2} holds true. However, this estimation can be inaccurate. The undesirable byproducts generated by the correlation between the signal and noise vectors in \eqref{1} cannot be ignored \cite{37}. To tackle this problem, diagonal loading (DL) technique can be used \cite{48,49}. Then the the following estimate of the auto-correlation matrix is used
\begin{equation}
	{\bf \hat R}_{\rm x} = \frac{1}{T_{\rm s}} \sum_{t = 1}^{T_{\rm s}} {\bf x}(t){\bf x}^H(t) + {\tau} {\bf I}_N,
\end{equation}
where ${\tau}$ is a small constant. Hence, the subspace leakage problem caused by the lack of snapshots can be mitigated. The estimation of ${\bf \hat R}_{\rm z}$ is analogous. 

It is also possible that the number of sources $K$ is unknown. Under this circumstance, the number of dominant singular values after truncated SVD can be approximately regarded as an acceptable estimate of the number of sources \cite{36,41}. There are also many other approaches that can be used to determine the number of sources. For example, the Bayesian approach has been introduced to achieve an automatic channel model complexity control and/or source enumeration in \cite{42,43,44}, which can be adopted for automatic estimation of the number of sources for {   the} L-shaped nested array as well.

\section{Simulation Results}\label{sec5}
In this section, several simulation examples are presented in order to evaluate the performance of the proposed iterative 2-D DOA estimation method. Throughout the simulations, an L-shaped nested array that consists of two nested subarrays with $N = 6$ elements along x-axis and z-axis is considered. For each nested subarray, the inner ULA consists of $N/2 = 3$ elements with spacing $d = \lambda/2$, while the outer ULA consists of the other three elements with spacing $2\lambda$. {  Hence, we have $\{\xi_n\}_{n=1}^N = \{1,2,3,4,8,12\}$.} The sources are modeled as random Gaussian processes and the noise is assumed to be spatially and temporally white. Consequently, the nested array forms a difference co-array with DOF $2S-1 = 23$. The optimal number of elements for each subarray is $Q \approx 10$, which is different from the conventional nested array ($Q = M = S = 12$) \cite{21}. In our examples, we assume that $K = 3$ sources are impinging on the L-shaped nested array from distinct directions
${(\theta_k,\phi_k)} \in \left\{(13^\circ,45^\circ), (25^\circ,29^\circ), (41^\circ,12^\circ)\right\}$, and {  $T_{\rm s} = 6400$}. The number of Monte Carlo trials is $P = 500$, while the scaling (reliability) factor in the proposed algorithm is $\mu = 0.95$. The designed tensor model \eqref{13} is used. The SNR is computed as
\begin{equation} \label{SNR}
    {\rm SNR}~[{\rm dB}] \triangleq 10 \log \frac{\left\| {\bf \tilde{T}}_{(2)} \right\|_{\rm F}^2} {\left\| {\tau {\bf{N}}} \right\|_{\rm F}^2} .
\end{equation}

Conventional L-shaped array-based DOA estimation approaches such as the JSVD \cite{12}, CESA \cite{15}, CSAP \cite{18}, TALA \cite{19}, SS \cite{22} and CPD \cite{28} are introduced for comparison. Note that the JSVD, CSAP, TALA and CPD can achieve an automatic pair-matching, while the CESA, SS and the proposed method require pairing of the elevation and azimuth angles. In the proposed algorithm, the initial DOA estimation result before mitigating the cross term is named as Step~1 while the final DOA estimation result after eliminating the cross term is referred to as Step~2.
\begin{figure}
	\centerline{\includegraphics[width=\columnwidth]{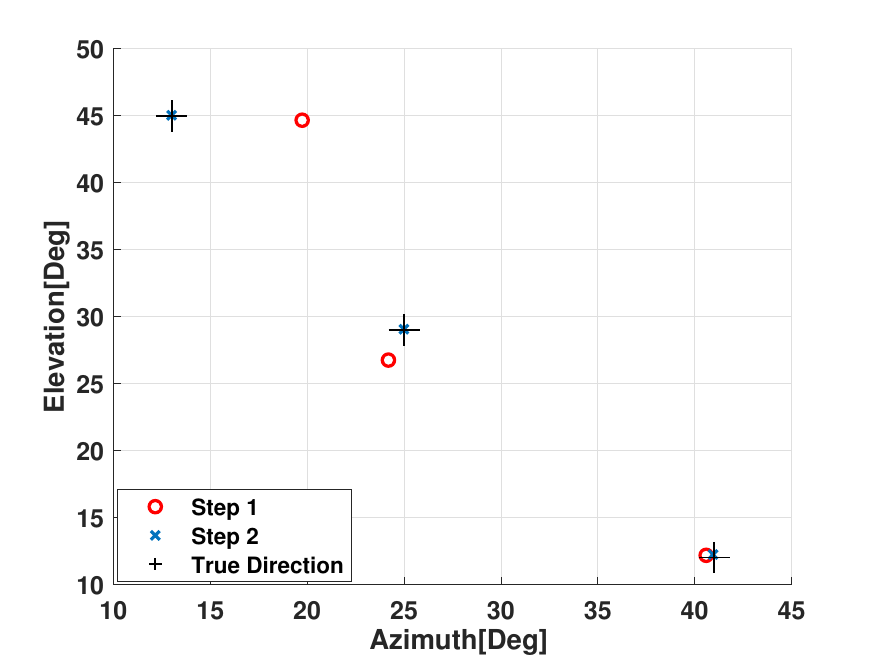}}
	\caption{2-D DOA estimation results for the proposed approach, $K = 3$, $L = 3$, SNR = 0~dB. The cross term mitigation improves the DOA estimation.}
	\label{fig1}
\end{figure}

\begin{figure}
	\centerline{\includegraphics[width=\columnwidth]{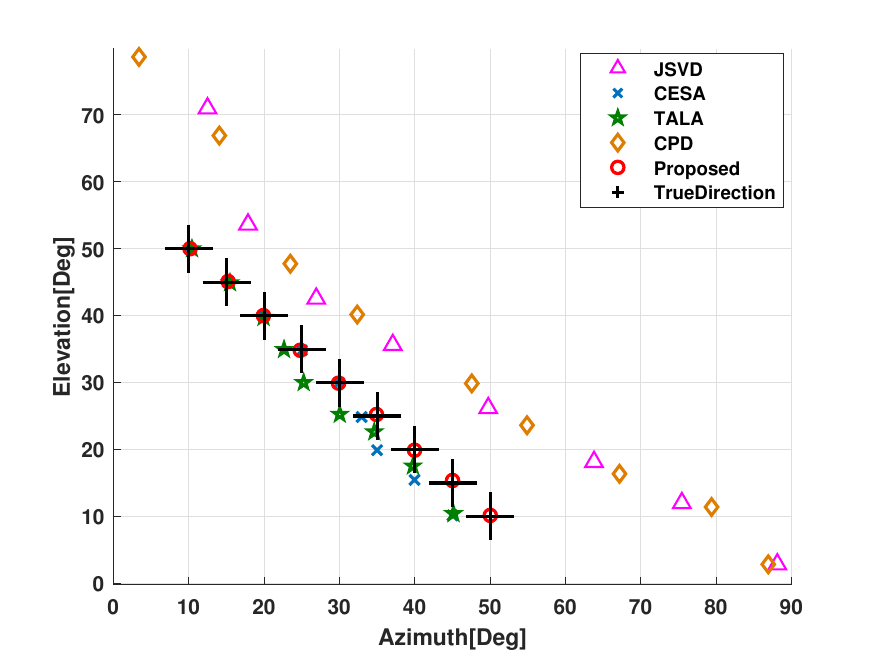}}
	\caption{A total of 12 elements (6 elements on each axis) are used to resolve 9 sources with 13~dB input SNR. The result is averaged over 500 trials.}
	\label{fig3}
\end{figure}

\subsection{Example~1: Effect of the Proposed Iterative 2-D DOA Estimation Method}
In our first example, the DOA estimates before and after cross term mitigation are shown to demonstrate the validity of the proposed method. 

In Fig.~\ref{fig1}, the DOA estimates for three sources obtained by the proposed algorithm with $L = 3$ iterations are shown. The SNRs are 0~dB. For Step~1, the DOA estimation is based on the initial received signal matrix ${\bf T}_{(2)}$ where cross term ${\bf T}_{{\rm ct},(2)}$ is present. Note that DOAs of all three sources are not correctly estimated. However, using the proposed iterative estimation method in \textbf{Algorithm~\ref{algorithm}}, it is possible to gradually eliminate the cross term. It can be observed that the DOA estimates in Step~2 are more accurate after a proper mitigation of the cross term. All three sources are resolved successfully.

{  Then, we assume that totally $K = 9$ sources are impinging on the L-shaped nested array with $\theta_k = 5 + 5k$, $\phi_k = 55 - 5k$, and $ k = 1,2,\cdots, K$. We also assume that the powers of all sources are identical, that is, we let the source SNRs all be 13~dB. The maximum number of iterations of the proposed algorithm is $L = 20$. Only the results of the DOA estimation after Step~2 (final estimates) by the proposed approach are shown. The DOA estimates obtained by the JSVD, CESA, TALA and CPD are also given. The other settings are the same as in the previous example. 

It can be observed in Fig.~\ref{fig3} that all source DOAs are estimated and paired by our proposed method correctly. The approaches used for comparison, however, fail to resolve the $K$ sources and show different levels of estimation error. It is also worth noting that some estimates by the proposed approach fall out of the grid slightly, which is caused by the cross term residue in Step~2. With the increase of the number of sources, the precise estimation of ${\bf T}_{{\rm ct},(2)}$ becomes more difficult as well as the determination of a proper reliability factor $\mu$ becomes increasingly difficult. Under this circumstance, the cross term is unavoidable and the DOA estimation accuracy can therefore degrade.}

\subsection{Example~2: Parameters Selection of the Proposed Algorithm}
\begin{figure}
	\centerline{\includegraphics[width=\columnwidth]{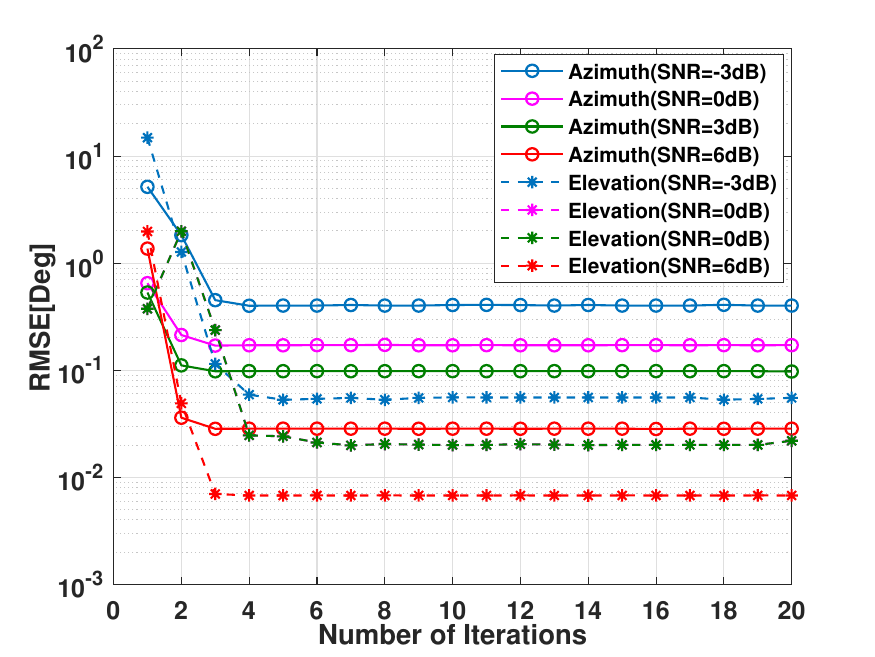}}
	\caption{RMSEs versus the number of iterations, 3 sources and 500 trials. At lower input SNR, more iterations are required to fully remove the cross term.}
	\label{fig2}
\end{figure}

\begin{figure}
	\centerline{\includegraphics[width=\columnwidth]{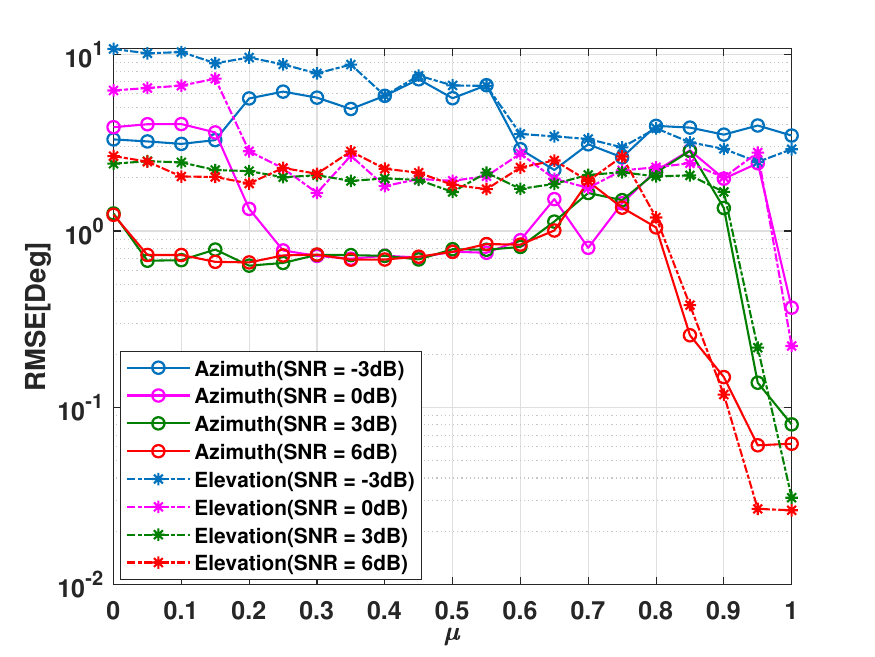}}
	\caption{RMSEs versus the reliability factor $\mu$, 3 sources, 500 trials. The lowest RMSE gives the optimal $\mu$ approximately.}
	\label{fig4}
\end{figure}

Here, we first aim at studying how many iterations are required for the proposed DOA estimation algorithm. The number of sources is $K = 3$. Assume that the maximum number of iterations is $L = 20$ and evaluate the root mean square error (RMSE) versus the number of iterations for several values of SNR. The azimuth RMSE is computed by
\begin{equation} \label{RMSE}
{\rm RMSE} = \sqrt {\frac{1}{{2PK}}\sum\limits_{k = 1}^K {\sum\limits_{p = 1}^P {{{\left( {{{\hat \theta }_k}(p) - {\theta _k}(p)} \right)}^2}} } },
\end{equation}
while the elevation RMSE is obtained similar by replacing $\theta_k$ with $\phi_k$. It can be seen in Fig.~\ref{fig2} that the proposed method converges after several iterations for both azimuth and elevation estimates. Note that the number of iterations required by the proposed method increases gradually with the decrease of the SNR, since the estimation of ${\bf T}_{{\rm ct},(2)}$ is more reliable at high SNR. The number of iterations $L$ required for convergence is no more than 4 when the SNR is above 0~dB. In some cases, only one iteration is sufficient.

Note also that if SNR is too low, the first DOA estimation results obtained in Step~1 of the proposed algorithm barely contains any sources spatial information. Under this circumstance, some other DOA estimation methods that perform well at low SNR can be used to initialize the proposed iterative algorithm. {   In fact, the performance of the k-means clustering algorithm used to find the unique eigenvalues may also degrade at low SNR. This is because outliers can appear when SNR is low, which may lead to an offset between the clustering centers and the corresponding eigenvalues. 
}

Next, we evaluate the optimal value of the reliability factor $\mu$ and show the RMSE performance of the proposed algorithm versus different values of $\mu$. Four cases with different SNRs (-3~dB, 0~dB, 3~dB and 6~dB) are considered. Although it is unnecessary when SNR is large enough, the maximum number of iterations is still set as $L = 20$ for each case. The reliability factor $\mu$ varies from zero to one with a fixed step size~0.05, while other parameters are unchanged as compared to the previous example.

It can be seen in Fig.~\ref{fig4} that the elevation and azimuth RMSEs are poor when SNR is -3~dB. This is because the signal component is interfered by the noise, and the reconstructed cross term ${\bf T}_{{\rm ct},(2)}$ in Step~2 of the proposed iterative algorithm contains no sources information but noise. Thus, the accuracy of the proposed iterative 2-D DOA estimation method is limited by the accuracy of the initial Step~1 when SNR is low, and as mentioned above, some other DOA estimation method that perform well at low SNR should be used to initialize the proposed iterative algorithm. From the other three cases, it can be seen that the final RMSEs for both elevation and azimuth estimates remain at a relatively high level when $\mu$ raises from zero to~0.8.
A turning point can be observed when $\mu >0.8$, after which the RMSEs for  both elevation and azimuth estimates decline rapidly. It can also be found that the decrease of RMSEs become insignificant once $\mu >0.95$. Thus, we can set $\mu = 0.95$ as the suboptimal value during our simulations. Although deriving the closed-form expression for computing the optimal reliability factor is not feasible, we use this example to demonstrate that $\mu$ can be determined in practice, and the algorithm is not very sensitive if a suboptimal value is selected. Indeed, the proposed method is valid with a suboptimal $\mu$. The only possible price is that a suboptimal scaling (reliability) factor may degrade the convergence speed, which means that more iterations may be required.

\subsection{Example~3: RMSE Performance versus SNR}
\begin{figure}
	\centering
	\subfloat[Elevation RMSE versus SNR]{%
		\includegraphics[width=\columnwidth]{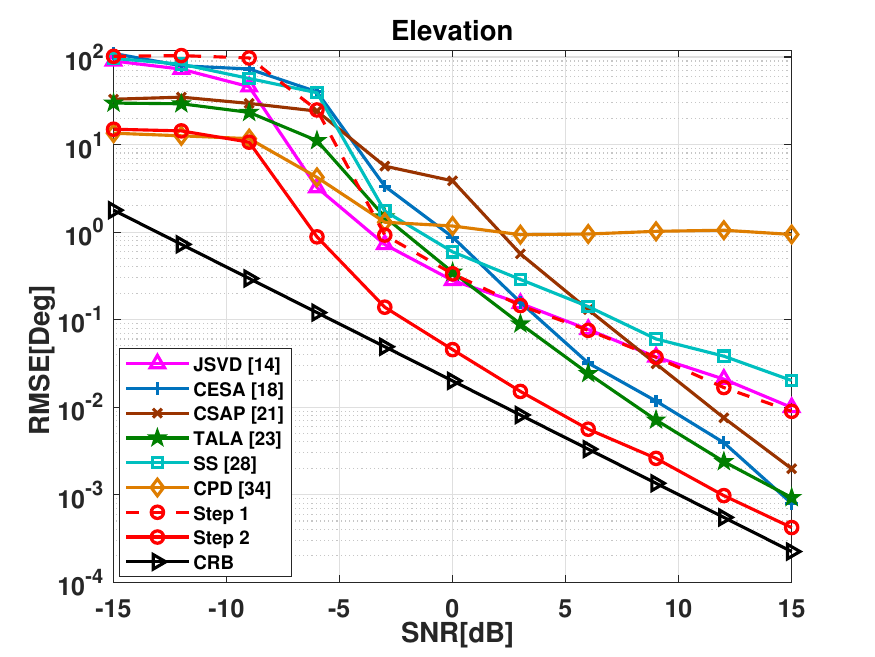}}
	\\
	\subfloat[Azimuth RMSE versus SNR]{%
		\includegraphics[width=\columnwidth]{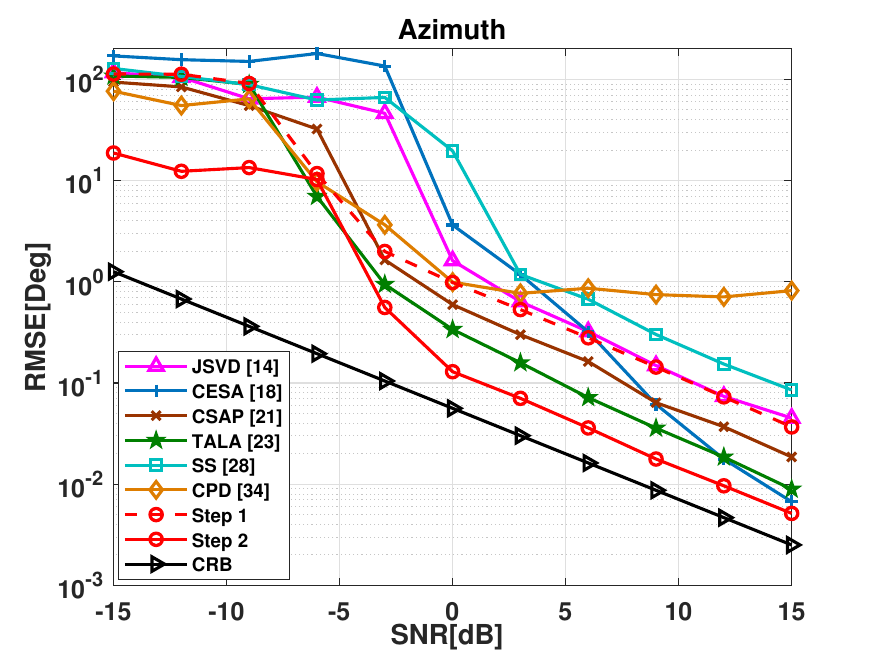}}
	\caption{RMSE versus SNR, 3 sources and 500 trials. The proposed approach obtains the best estimation accuracy due to the elimination of the cross-term and the present tensor structure exploitation.}\label{fig5}
\end{figure}

Our third example aims to illustrate the DOA estimation performance of the proposed iterative algorithm in terms of RMSE. Three sources are placed at ${(\theta_k,\phi_k)} \in \left\{(10^\circ,45^\circ), (20^\circ,40^\circ), (30^\circ,35^\circ)\right\}$. To ensure the validity of the proposed algorithm, the maximum number of iterations is set as $L = 20$ and $\mu = 0.95$. The JSVD, CESA, CSAP and SS methods exploit the signal covariance matrix model. The CPD algorithm uses the higher-order tensor model \eqref{13}, while the TALA and the proposed approaches are based on the reshaped tensor model \eqref{15}. The CRB of the proposed method is also presented. The other settings are unchanged.

The elevation and azimuth RMSEs of the algorithms tested are shown in Fig.~\ref{fig5}. {  It can be seen that the conventional CPD method suffers from the convergence problem of the ALS algorithm, especially for a higher-order tensor. The estimation accuracy for elevation and azimuth angles are quite poor. The SS method can entirely eliminate the cross term and the signal covariance matrix is positive semidefinite for any finite number of snapshots. However, this method averages the signal covariance matrices of all subarrays in the co-array domain on both x-axis and z-axis. The multi-dimensional structure between those subarrays is ignored. Thus, the corresponding RMSEs are relatively poor. The JSVD approach takes advantage of the CCM and improves the estimation accuracy slightly, while the CSAP method exploits the conjugate symmetric property of the array manifolds to increase both the array aperture and the number of snapshots to achieve lower estimation error. Both methods require SVD. The CESA method that deals with the first column, the first row and diagonal entries of the CCM is also used. It provides a good estimation accuracy. However, it requires to perform spectrum search twice. Moreover, additional computations are required for pairing of azimuth and elevation angles. To exploit the multi-dimensional structure of the subarrays in the co-array domain, the TALA method is also utilized. Because it ignores the Vandermonde structure of the factor matrices and the influence of the cross term, the performance improvement appears to be insignificant as compared to that of the CESA algorithm. In the proposed approach, the DOA estimates in Step~1 are also relatively poor because the cross term degrades the performance significantly. It can be shown that the TALA algorithm that uses the same tensor model improves the RMSEs as compared to the results of Step~1 of the proposed algorithm. It is because the TALA algorithm deals with the designed higher-order tensor directly. The cross term in tensor form is a sparse tensor, which has less influence on the DOA estimation results obtained by tensor decomposition. However, the use of the TALA algorithm demands much more computational resources, and the method is unstable especially when the number of targets is unknown.} After the cross term mitigation, the proposed method surpasses the other methods and shows the lowest RMSE threshold. It is because it exploits the multi-dimensional structure of the received signal for all subarrays and removes the cross term efficiently. A computationally efficient tensor decomposition method is used to conduct 2-D DOA estimation, while the proposed iterative method is used to remove the cross term.

\subsection{Example~4: Probability of Resolution versus SNR}
Finally, we evaluate the methods tested in terms of the probability of resolution for two closely spaced sources. We assume only two sources in this example at ${(\theta_k,\phi_k)} \in \left\{(15^\circ,30^\circ), (16^\circ,31^\circ)\right\}$. The other parameters are unchanged as compared to the previous example. These two sources are considered to be resolved if
\begin{equation}
\begin{aligned}
  & \left\| {{{\hat \theta }_k} - {\theta _k}} \right\| \le \left\| {{\theta _1} - {\theta _2}} \right\|/2 \\
  & \left\| {{{\hat \phi }_k} - {\phi _k}} \right\| \le \left\| {{\phi _1} - {\phi _2}} \right\|/2, \quad k = 1,2
\end{aligned}
\end{equation}
holds true.

\begin{figure}
	\centering
	\subfloat[Elevation resolution versus SNR]{%
		\includegraphics[width=\columnwidth]{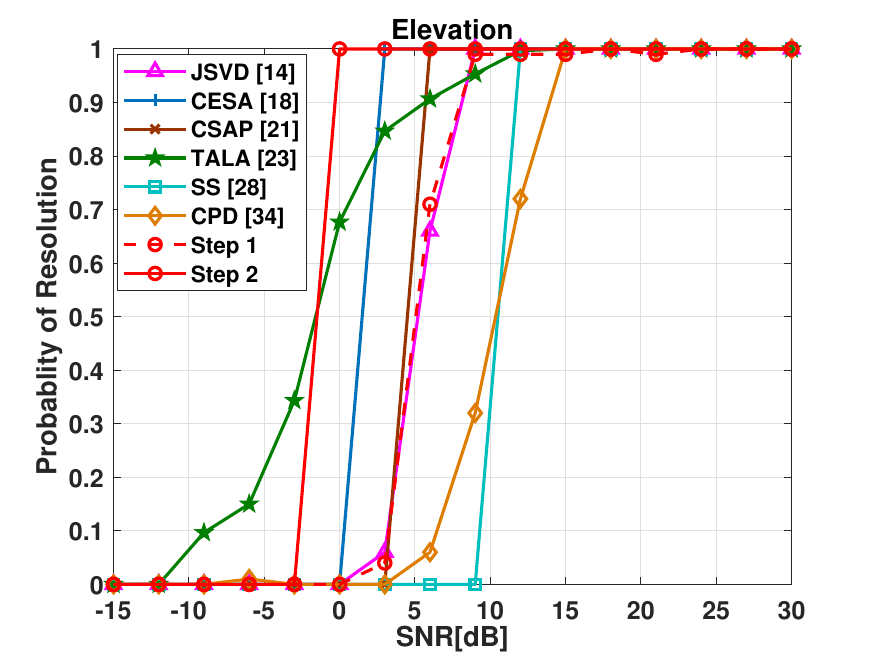}}
	\\
	\subfloat[Azimuth resolution versus SNR]{%
		\includegraphics[width=\columnwidth]{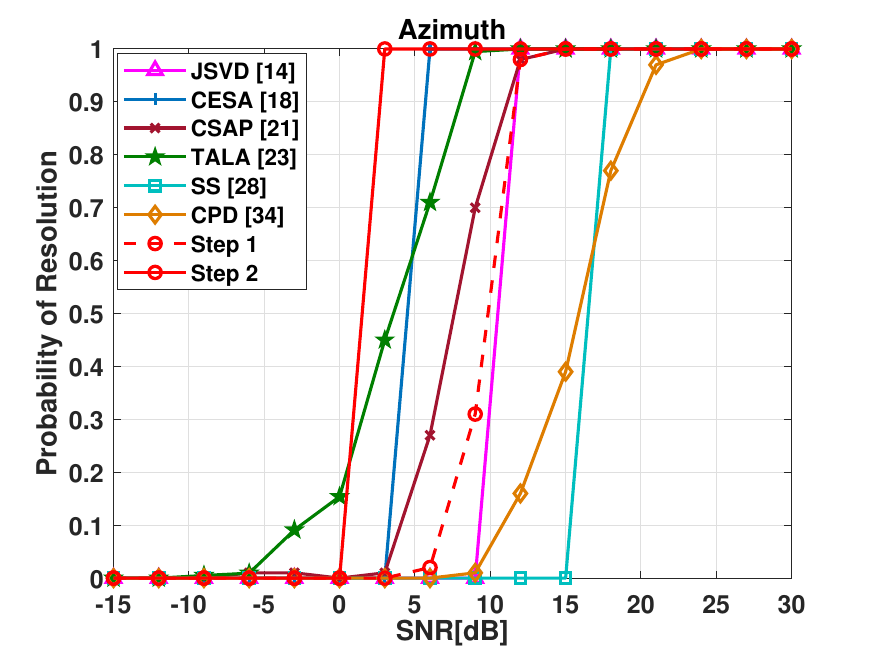}}
	\caption{Probability of resolution versus SNR, two closely spaced sources and 500 trials. The proposed approach surpasses other methods by achieving the lowest detection threshold.}\label{fig6}
\end{figure}

It can be seen in Fig.~\ref{fig6} that all methods tested achieve perfect resolution at high SNR. For each method, the elevation resolution threshold is substantially smaller than its counterpart in azimuth, which is reasonable since the RMSEs of elevation estimation are usually better than those of azimuth estimation as shown in previous examples. {  The performance of the CPD as well as SS methods is poor, while the JSVD, CSAP and CESA approaches improve the resolution performance at different levels. The DOA estimates in Step~1 of the proposed method are unsatisfactory due to the influence of the cross term. It is possible for the TALA algorithm to use the designed tensor model in order to resolve two closely spaced sources at very low SNR. The threshold, however, is worse than that of the proposed method due to the destructive influence of the cross term. Indeed, the cross term is masked by the noise when SNR is low, and the TALA algorithm can capture only the structure of the signal term, while the structure of the cross term is different and cannot be captured by the TALA algorithm. In fact, the cross term for the TALA method has the same effect as an additional noise component, and the cross term degrades the performance of the TALA method even when SNR is high. With the increase of SNR, the DOA estimation accuracy improvement provided by the TALA algorithm degrades since the cross term starts to dominate the noise term. This is the main reason behind the fact that the curves (for azimuth and elevation) for the TALA algorithm in Fig.~\ref{fig6} are relatively flat compared to the curves for the other methods tested (which is also true for the CPD methods).} Consequently, the proposed iterative 2-D DOA estimation method enables the lowest threshold for both elevation and azimuth resolution and, hence, achieves a better resolution performance as it can effectively mitigate the cross term.

\section{Conclusion}\label{sec6}
An iterative 2-D DOA estimation method via tensor modeling has been proposed for {  the} L-shaped nested array. In the proposed method, a higher-order tensor has been designed to exploit the multi-dimensional structure of the received signal for all subarrays in co-array domain. The designed tensor model improves the {  system's DOF} by optimizing the number of subarrays for SS technique. A computationally efficient tensor decomposition method has been then developed to decompose the Vandermonde factor matrices, whose vectors of generators provide the { sources' spatial information}. The cross term caused by the correlated signal and noise components of the received signal in co-array domain is estimated and removed in the second step of our methods based on the DOA estimates obtained at the first step, and then steps are repeated iteratively to achieve a better DOA estimation performance. Therefore, the received signal can be modified gradually during iterations. Comparing with existing DOA estimation methods for {  the} L-shaped nested array, the proposed method can take advantage of the multi-dimensional structure of the received signal, it is also capable of mitigating the cross term. The parameter identifiability of the designed tensor model has been significantly improved. Simulation results have verified that the proposed method achieves a better accuracy and higher resolution in the problem of 2-D DOA estimation for {  the} L-shaped nested array as compared to existing techniques.

\balance
\bibliographystyle{IEEEtran}
\bibliography{IEEEabrv,Ref}

\end{document}